\documentclass[aps,prl,preprintnumbers,amsmath,amssymb,latexsym,nofootinbib,array,enumerate,letter,twocolumn,superscriptaddress]{revtex4-1}
\usepackage{subfigure}
\usepackage{cancel}
\usepackage{here}
\usepackage{comment,braket}
\usepackage{epsf}
\usepackage{amsmath}
\usepackage{graphics}
\usepackage{amsfonts}
\usepackage{amssymb}
\usepackage{latexsym}
\usepackage{color}
\usepackage{ulem}
\input{colordvi.tex}
\usepackage{feynmp}
\usepackage{caption}
\usepackage[pdftex]{graphicx}
\usepackage{bm}
\usepackage[pdftex]{hyperref}
\usepackage[svgnames]{xcolor}
\definecolor{phthaloblue}{rgb}{0.0, 0.06, 0.54}
\hypersetup{
    colorlinks=true,
    linkcolor=phthaloblue,
    citecolor=blue,
    filecolor=blue,
    urlcolor=phthaloblue,
    }

\newcommand{\beq}{\begin{eqnarray}} 
\newcommand{\eeq}{\end{eqnarray}}

\def\({\left(}
\def\){\right)}
\def\[{\left[}
\def\]{\right]}

\newcommand{\bel}[1] {\begin{equation}\label{#1}}
\newcommand{\beal}[1] {\begin{eqnarray}\label{#1}}
\newcommand{\be}{\begin{equation}}
\newcommand{\ee}{\end{equation}}
\newcommand{\bea}{\begin{array}} 
\newcommand{\eea}{\end{array}}

\begin{document}
%\preprint{RESCEU-1/16}

\title{Affleck-Dine Leptogenesis from Higgs Inflation}

\author{Neil D. Barrie}
\email{nlbarrie@ibs.re.kr}
\affiliation{Center for Theoretical Physics of the Universe, Institute for Basic Science (IBS),
Daejeon, 34126, Korea}
\preprint{CTPU-PTC-21-21}
	
\author{Chengcheng Han}
\email{hanchch@mail.sysu.edu.cn}
\affiliation{School of Physics, Sun Yat-Sen University, Guangzhou 510275, China}

\author{Hitoshi Murayama}
\email{hitoshi@berkeley.edu, hitoshi.murayama@ipmu.jp, Hamamatsu Professor}
\affiliation{Department of Physics, University of California, Berkeley, CA 94720, USA}
\affiliation{Kavli Institute for the Physics and Mathematics of the
  Universe (WPI), University of Tokyo,
  Kashiwa 277-8583, Japan}
\affiliation{Ernest Orlando Lawrence Berkeley National Laboratory, Berkeley, CA 94720, USA}

\date{\today}

\begin{abstract}
We find that the triplet Higgs of the Type II seesaw mechanism can simultaneously generate the neutrino masses and observed baryon asymmetry, while playing a role in inflation. We survey the allowed parameter space and determine that this is possible for triplet masses as low as a TeV, with a preference for a small vacuum expectation value for the triplet $v_\Delta < 10 $ keV. This requires that the triplet Higgs must decay dominantly into the leptonic channel. Additionally, this model will be probed at the future 100 TeV collider, upcoming lepton flavor violation experiments such as Mu3e, and neutrinoless double beta decay experiments. Thus, this simple framework provides a unified solution to the three major unknowns of modern physics - inflation, the neutrino masses, and the observed baryon asymmetry - while simultaneously providing unique phenomenological predictions that will be probed terrestrially at upcoming experiments.

%We investigate the possibility of simultaneously explaining inflation, the neutrino masses and the baryon asymmetry through extending the Standard Model by the triplet Higgs of the Type II seesaw mechanism. The neutrino masses are generated by the vacuum expectation value of the triplet Higgs, while a combination of the triplet and doublet Higgs' plays the role of the inflaton. Additionally, the dynamics of the triplet, and its inherent lepton number violating interactions, lead to the generation of a lepton asymmetry during inflation via the Affleck-Dine mechanism. The resultant baryon asymmetry, inflationary predictions and neutrino masses from this simple Standard Model extension are consistent with current observational and experimental results. 
\end{abstract}

%\preprint{}

\maketitle

{\bf Introduction.}---%
Despite the great successes of the Standard Model (SM) at describing low energy scales, there remain many open problems that demand the existence of new physics. These issues include  the mechanism for the epoch of rapid expansion in the early universe (inflation~\cite{Brout:1977ix, Sato:1980yn, Guth:1980zm, Linde:1981mu, Albrecht:1982mp}), the origin of neutrino masses, and the source of the observed baryon asymmetry. Each of these mysteries is tied to early universe physics, with any associated discoveries having significant implications for both particle physics and cosmology. 

% Although inflation is currently considered an integral component of standard cosmology, the exact nature of the inflationary epoch remains unknown. The existence of non-zero neutrino masses and the observed baryon asymmetry are key pieces of evidence for physics beyond the SM.

An exciting possibility to explore is whether each of these unknowns could be explained within a simple unified framework. There have been multiple attempts to do so in the past, but it is difficult to provide solutions to all three problems simultaneously with a single addition to the SM. For example, the SM plus three right-handed neutrinos can explain the neutrino masses via the seesaw mechanism~\cite{Minkowski:1977sc, Yanagida:1979as, Glashow:1979nm, GellMann:1980vs} and generate the baryon asymmetry through Leptogenesis~\cite{Fukugita:1986hr}, but not the inflationary sector.  Inflationary Baryogenesis has been widely investigated in the  literature~\cite{Affleck:1984fy,Hertzberg:2013mba, Lozanov:2014zfa, Yamada:2015xyr, Bamba:2016vjs,Bamba:2018bwl, Cline:2019fxx,Barrie:2020hiu, Lin:2020lmr, Kawasaki:2020xyf, Kusenko:2014lra, Wu:2019ohx, Charng:2008ke, Ferreira:2017ynu, Rodrigues:2020dod, Lee:2020yaj, Enomoto:2020lpf}, but with few cases able to simultaneously explain each of the issues named above.\footnote{Inflation with Leptogenesis by the right-handed sneutrino has been considered in ~\cite{Murayama:1992ua, Murayama:1993xu}.} We present a model herein that represents a simple and well-motivated realization of this idea.

In this letter, we study the possibility that these three problems can be solved through the simple extension of the SM by the triplet Higgs of the Type II Seesaw mechanism. It has been known for a long time that with the addition of one triplet Higgs the baryon asymmetry cannot be generated through thermal Leptogenesis, but rather requires the introduction of a second triplet Higgs ~\cite{Ma:1998dx, Hambye:2000ui, DAmbrosio:2004rko, Chun:2005ms, Chun:2006sp}, or a right-handed neutrino ~\cite{Hambye:2003ka, ODonnell:1993obr, Guo:2004mp, Antusch:2004xy, Gu:2006wj}.  In Ref. \cite{Senami:2001qn},  a mechanism for  non-thermal Leptogenesis was proposed involving the Affleck-Dine mechanism, but involved the addition of two triplet Higgses to the framework of  supersymmetry. However, to explain the existence of neutrino masses, only one triplet Higgs is required. We propose a mechanism by which successful Leptogenesis and neutrino mass generation can occur, with the addition of only a single triple Higgs. The triplet Higgs, in combination with the SM Higgs, will simultaneously give rise to a Starobinsky-like inflationary epoch \cite{Starobinsky:1980te,Whitt:1984pd,Jakubiec:1988ef,Maeda:1988ab,Barrow:1988xh,Faulkner:2006ub,Bezrukov:2011gp}. This provides a unique connection between the high energy dynamics of the early Universe and those at terrestrial colliders, which give novel phenomenological predictions that will be probed at future experiments. 
\vspace{2pt}

%Following, we consider a complex field $\phi$ with a global $U(1)$ charge $Q$ and non-trivial potential $ V $. 
{\bf Baryogenesis from a Complex Inflaton }--- 
The fundamental feature of the Affleck-Dine mechanism is the generation of angular motion in the phase of a complex scalar field $\phi$ that is charged under a global $U(1)$ symmetry \cite{Affleck:1984fy}. Assuming it acquires a large initial field value in the early universe, $ \phi $ will begin to oscillate once the Hubble parameter becomes smaller than its mass $m$.  If the scalar potential $ V $ contains an explicit $U(1)$ breaking term, a net $U(1)$ charge asymmetry will be generated by this motion.  The asymmetry number density associated with the $U(1)$ charge is then given by,
\begin{eqnarray}
n_Q = 2 Q \textrm{Im}[\phi^\dagger \dot \phi]=  Q \chi^2 \dot{\theta }~,
\end{eqnarray}
where  $\phi= \frac{1}{\sqrt{2}}\chi e^{i\theta}$. Therefore, in order to obtain a non-zero $n_Q$, we require non-zero vacuum value for $\chi$ and the motion of the complex phase $\theta$. 
This is easily realized if the  $\phi$ field also plays the role of the inflaton, with an initial non-vanishing $\chi$ and $\theta$. For a general potential for the $\phi$ field, we can separate the $U(1)$ conserving and non-conserving components,
\begin{eqnarray}
U(\phi) \equiv  U(\chi,\theta) = U_c(\chi) + U_{b}(\chi,\theta)~,
\end{eqnarray}
where $U_c(\chi)$ contains the $U(1)$ conserving terms which we assume dominate the potential during inflation, and $U_{b}(\chi,\theta)$ represents the $U(1)$ breaking terms. %The simplest assumption is that $U_1(|\phi|) = \frac{1}{2}m^2 |\phi|^2$. However, such simple model is already excluded by current observations from CMB. 
If the kinetic term of $\phi$ is canonically normalized, then the Lagrangian can be written as,
\begin{eqnarray}
{\mathcal L} =  - \frac{1}{2}  g^{\mu\nu} \partial_\mu \chi \partial_\nu \chi 
 -\frac{1}{2} f(\chi)  g^{\mu\nu} \partial_\mu \theta \partial_\nu \theta- U(\chi,\theta) ~,
\end{eqnarray}
where $f(\chi)=\chi^2$. Then the equations of motion for $\chi$ and $\theta$ are as follows,
\begin{eqnarray}
&& \ddot{\chi} -\frac{1}{2} f'(\chi)  \dot \theta^2 + 3 H \dot \chi + U_{,\chi} =0  ~,\nonumber \\
&& \ddot{\theta} + \frac{f'(\chi)}{f(\chi)}  \dot \theta \dot \chi + 3 H \dot \theta +\frac{1}{f(\chi)} U_{,\theta} =0 ~.
\label{thetaEOM}
\end{eqnarray} 
Assuming during inflation, both $\chi$ and $\theta$ are slow rolling,
\begin{eqnarray}
\dot{\chi} \simeq -\frac{M_p U_{,\chi}}{\sqrt{3U}}, ~~\textrm{and}~~ \dot{\theta} \simeq -\frac{M_p U_{,\theta}}{f(\chi)\sqrt{3U}} ~.
\label{dottheta_inf11}
\end{eqnarray}
From this we may estimate the Q-number density at the end of inflation,
\begin{eqnarray}
n_Q \approx -Q \chi^2_{\rm end }  \frac{M_p U_{,\theta}}{f(\chi_{\rm end })\sqrt{3U_{\rm end}}}.
\end{eqnarray}

%After inflation, the $\chi$ field would oscillate around the minimum of the potential which can be approximated as a quadratic potential. At this
%stage the universe is matter-like and the the amplitude of $\chi$ decrease with a expansion factor $a$. 
%From the Eq. \ref{thetaEOM}, we can get
%\begin{eqnarray}
%\frac{d ( a^3 n_Q)}{dt}  = Q a^3 U_{,\theta}
%\label{dlepton}
%\end{eqnarray}
%Thus if the $U(1)$ breaking term decrease faster than $a^3$, the $a^3 n_Q$ becomes conserved. Then the $n_Q$ is just redshifted by a factor $a^3$ after inflation.

Consequently, if the $U(1)$ symmetry is composed of the global $U(1)_B$ or $U(1)_L$ symmetries, a baryon asymmetry can be generated prior to the Electroweak Phase Transition. In the following we will show that the $\phi$ field can be a mixed state of the SM and triplet Higgs', with a complex phase associated with the $U(1)_L$ symmetry.
\vspace{2pt}

{\bf Model Framework}--- 
 We now introduce the Lagrangian describing the SM Higgs doublet $ H $, and the triplet Higgs $\Delta$. The scalars are parameterized by,
\begin{eqnarray}
H =\left(
\begin{array}{c}
 ~~ h^{+}      \\
  h     
\end{array}
\right)
,~~
\Delta =\left(
\begin{array}{cc}
  \Delta^+/\sqrt{2} & \Delta^{++}      \\
  \Delta^0& -\Delta^+/\sqrt{2}     
\end{array}
\right)~,
\end{eqnarray}
where $h$ and $\Delta^0$ are the neutral components of $H$  and $\Delta$ respectively. The  $\mathcal L_{\textrm{Yukawa}}$ term contains not only the Yukawa interactions of the SM fermions, but also a new interaction between the left-handed leptons and the triplet Higgs $\Delta$, 
\begin{eqnarray}
\mathcal L_{Yukawa} = \mathcal L^{\rm SM}_{\textrm{Yukawa}}-\frac{1}{2}y_{ij} \bar L^c_i \Delta L_j + h.c. 
\label{nu_interaction}
\end{eqnarray}
This interaction term will generate the neutrino mass matrix, once $\Delta^0$ obtains a non-zero VEV. Through this interaction we assign a lepton charge of $Q_L=-2$ to the triplet Higgs, thus opening the possibility for it to play a role in the origin of the baryon asymmetry. 

The potential for the neutral Higgs' components is,
\begin{eqnarray}
V(h, \Delta^0) &&= -m_H^2 |h|^2+ m_\Delta^2 |\Delta^0|^2 + \lambda_H  |h|^4 + {\lambda_\Delta}  |\Delta^0|^4  \nonumber \\
&& + {\lambda_{H\Delta}} |h|^2 |\Delta^0|^2 + \left(\mu h^2 {\Delta^0}^* + \frac{\lambda_5}{M_p} |h|^2  h^2  {\Delta^0}^* \right. \nonumber \\
&& \left. + \frac{\lambda^\prime_5}{M_p} |\Delta^0|^2 h^2  {\Delta^0}^* +h.c. \right) +...  ~,
\end{eqnarray}
%$ \lambda_\Delta = \lambda_2+\lambda_3~, $ and  ~$ \lambda_{H\Delta} = \lambda_1+\lambda_4 $, and 

Importantly for our model, the necessary $\mu $ coupling between the SM and triplet Higgs' inherently violates lepton number as defined through the triplet Higgs Yukawa interaction. Additionally, we have included dimension five lepton violating operators that are suppressed by $M_p$, since during inflation the field value is close to the Planck scale and as such they can dominate over the $\mu$ term. However, the higher dimensional terms will play no role in low energy physics.

%{\color{red}The LHC has placed a  lower limit on the mass term $m_\Delta \gtrsim800 $ GeV through searches for the associated doubly-charged Higgs.} 
The potential couplings are constrained by requiring the stability condition, and the non-vanishing $\Delta^0$ VEV can be approximated in the limit  $m_\Delta \gg v_{\textrm{EW}}$, $v_\Delta \equiv \langle \Delta^0 \rangle \simeq - \frac{\mu v_{\textrm{EW}}^2}{ 2 m^2_\Delta}$~,
where the SM Higgs VEV is $v_{\textrm{EW}} = 246$ GeV. The $\Delta^0$ VEV is bounded by $\mathcal{O}(1) \textrm{~GeV}>|\langle \Delta^0 \rangle| \gtrsim 0.05$ eV, in order to generate the observed neutrino masses, while ensuring $y_\nu$ is perturbative up to $ M_p$. The upper bound on the $\Delta^0$ VEV is derived from T-parameter constraints determined by precision measurements \cite{Kanemura:2012rs}.
%The LHC has placed the lower limit $m_\Delta \gtrsim 800 $ GeV from searches for the associated doubly-charged Higgs~\cite{ATLAS:2017xqs}.\\

%, such as in the usual Higgs inflation model
Although the above doublet-triplet Higgs model includes all the ingredients for generating the baryon asymmetry during inflation, the current data from CMB observations excludes their simple polynomial potential as the source of inflation \cite{Planck:2018jri}. One resolution to this problem is the addition of non-minimal couplings between the Higgs' and the Ricci scalar. Then full Lagrangian is,
\begin{eqnarray}
\frac{\mathcal L}{\sqrt{-g}} & & = -\frac{1}{2} M_p^2 R -F(H,\Delta) R -g^{\mu\nu} (D_\mu H)^\dagger (D_\nu H) \nonumber \\
&& -g^{\mu\nu} (D_\mu \Delta)^\dagger (D_\nu \Delta) -V(H, \Delta) + \mathcal L_{\textrm{Yukawa}}~,\label{Lagrange1}
\end{eqnarray}

\vspace{2pt}
{\bf Trajectory of Inflation}--- The inflationary setting will be induced by both Higgs' through their non-minimal couplings to gravity. These couplings act to flatten the scalar potential at large field values. This form of inflationary mechanism has been utilized in standard Higgs inflation, and results in a Starobinsky-like inflationary epoch \cite{Starobinsky:1980te,Bezrukov:2007ep,Bezrukov:2008ut,GarciaBellido:2008ab,Barbon:2009ya,Barvinsky:2009fy,Bezrukov:2009db,Giudice:2010ka,Bezrukov:2010jz,Burgess:2010zq,Lebedev:2011aq,Lee:2018esk,Choi:2019osi}. We consider the following non-minimal coupling, 
\begin{eqnarray}
F(H,\Delta) =  \xi_H |h|^2 + \xi_\Delta |\Delta^0|^2=\frac{1}{2}\xi_H \rho^2_H+\frac{1}{2} \xi_\Delta \rho^2_\Delta
\end{eqnarray}
where we have utilized the polar coordinate parametrization  $h \equiv \frac{1}{\sqrt{2}} \rho_{H} e^{i\eta}$,  $\Delta^0 \equiv \frac{1}{\sqrt{2}}\rho_{\Delta} e^{i\theta}$.
An inflationary framework consisting of two non-minimally coupled scalars  has been found to exhibit a unique inflationary trajectory \cite{Lebedev:2011aq}. In the large field limit the ratio of the two scalars is fixed,
\begin{eqnarray}
\frac{ \rho_{H}}{\rho_{\Delta}} \equiv \tan \alpha = \sqrt{\frac{2\lambda_\Delta \xi_H -\lambda_{H\Delta} \xi_\Delta}{ 2\lambda_H \xi_\Delta -\lambda_{H\Delta} \xi_H  }}~.
\end{eqnarray}
To ensure the evolution of this trajectory, we require $2\lambda_\Delta \xi_H -\lambda_{H\Delta} \xi_\Delta>0$ and $2\lambda_H \xi_\Delta -\lambda_{H\Delta} \xi_H>0$. The derivation of this trajectory is given in Supp.~IE. 
%To ensure $ \alpha $ variation does not play a role, we assume  $ \xi_H \sim \xi_\Delta $, fixing the field trajectory.
The inflaton can then be defined as $ \varphi $, through the relations,
\begin{eqnarray}
&&  \rho_{H} = \varphi \sin \alpha,~\rho_{\Delta} = \varphi \cos \alpha,~  \nonumber \\
&& \xi \equiv \xi_H \sin^2 \alpha + \xi_\Delta \cos^2 \alpha~.
\end{eqnarray}
The Lagrangian becomes,
\begin{eqnarray}
\frac{\mathcal L}{\sqrt{-g}} &=& -\frac{1}{2} M_p^2 R -\frac{1}{2} \xi  \varphi^2  R  - \frac{1}{2} g^{\mu\nu} \partial_\mu \varphi \partial_\nu \varphi  \nonumber \\
&& -\frac{1}{2} \varphi^2 \cos^2\alpha~ g^{\mu\nu} \partial_\mu \theta \partial_\nu \theta-V(\varphi, \theta) ~,
\label{lagrang1}
\end{eqnarray}
where 
\begin{eqnarray}
\hspace{-0.5cm}V(\varphi, \theta)  = \frac{1}{2} m^2 \varphi^2 + \frac{\lambda}{4} \varphi^4 + 2\varphi^3 \left(\tilde \mu   + \frac{\tilde \lambda_5}{M_p} \varphi^2\right) \cos\theta ,
\end{eqnarray}
and
\begin{eqnarray}
 && m^2 = m^2_{\Delta} \cos^2 \alpha  -m_H^2 \sin^2 \alpha~,  \nonumber \\
 && \lambda = \lambda_H \sin^4 \alpha + \lambda_{H\Delta} \sin^2 \alpha \cos^2 \alpha + \lambda_{\Delta} \cos^4 \alpha~,  \nonumber \\
 && \tilde \mu  = -\frac{1}{2\sqrt{2}}\mu \sin^2\alpha  \cos \alpha~,  \nonumber \\
 && \tilde \lambda_5  = - \frac{1}{4\sqrt{2}} (\lambda_5 \sin^4\alpha  \cos \alpha + \lambda^\prime_5 \sin^2\alpha  \cos^3 \alpha)~.
\end{eqnarray}

Since the generated lepton asymmetry is dependent upon the motion of $\theta$, we consider it to be a dynamical field. During inflation, $m \ll \varphi$, meaning that the quartic potential term dominates during the inflationary epoch. 

We translate the Lagrangian in Eq. (\ref{lagrang1}) from the Jordan frame to the Einstein frame, utilizing the transformations \cite{Wald:1984rg,Faraoni:1998qx}, $\tilde{g}_{\mu \nu} =\Omega^2 g_{\mu\nu},~~ \Omega^2= 1+ \xi \varphi^2/M^2_p~$, and reparametrizing $\varphi$ in terms of the canonically normalized scalar $\chi$. Obtaining the final Einstein frame Lagrangian,
\begin{eqnarray}
\frac{\mathcal L}{\sqrt{-g}} = -\frac{M_p^2}{2}  R   - \frac{1}{2}  g^{\mu\nu} \partial_\mu \chi \partial_\nu \chi \nonumber \\-
 \frac{1}{2} f(\chi)  g^{\mu\nu} \partial_\mu \theta \partial_\nu \theta- U(\chi,\theta) ~,
\end{eqnarray}
where 
\begin{eqnarray}
&& \hspace{-0.5cm}f(\chi) \equiv  \frac{\varphi(\chi)^2 \cos^2 \alpha}{\Omega^2(\chi)},  ~\textrm{and}~ U(\chi,\theta) \equiv  \frac{V(\varphi(\chi), \theta)}{\Omega^4(\chi)}~,
\label{potential}
\end{eqnarray}
with the $ \chi $ potential replicating the Starobinsky form in the large field limit,
\begin{eqnarray}
U_{\textrm{inf}}(\chi)= \frac{3}{4}m_S^2 M_p^2 \left(1-e^{-\sqrt{\frac{2}{3}}\frac{\chi}{Mp}}\right)^2~,
\end{eqnarray}
where $ m_S =\sqrt{\frac{\lambda M_p^2}{3\xi^2}}\simeq 3 \cdot 10^{13} $ GeV \cite{Faulkner:2006ub,Akrami:2018odb}. We will consider model parameters that ensure the inflationary trajectory is negligibly affected by the dynamics of $ \theta $. Under this assumption, the resultant inflationary observables are consistent with the Starobinsky model, and are in excellent agreement with current observational constraints \cite{Akrami:2018odb}. See Supp.~I for details of the inflationary epoch and observational predictions.
\vspace{2pt}

{\bf Lepton Number Density from the triplet Higgs}--- % 
%Now that we have detailed the inflationary dynamics, 
During inflation, we identify the inflaton  field $\phi$ as a mixed state of the doublet Higgs and triplet Higgs. The corresponding potential becomes, 
\begin{eqnarray}
U(\chi, \theta)=   \frac{m^2 \varphi^2(\chi)+ \lambda \varphi^4(\chi)  }{\Omega^4(\chi)}  +  \frac{2 \tilde \mu \varphi^3(\chi) + 2\frac{\tilde \lambda_5}{M_p}   \varphi^5(\chi)  }{\Omega^4(\chi)} \cos \theta \nonumber \\
\end{eqnarray}
and the lepton number density is  modified as,
\begin{eqnarray}
n_L=Q_L \varphi^2(\chi) \dot\theta \cos^2 \alpha    ~.
\end{eqnarray}
where $\alpha$ is the mixing angle between the doublet and triplet Higgs' during inflation.  
%The factor in the kinetic term becomes $f(\chi) = \frac{\varphi^2(\chi) \cos^2 \alpha}{\Omega^2(\chi)}$. Note that at the end of the inflation $\Omega \sim  \mathcal O(1)$. 
During inflation, the $\chi$ field approaches $M_{p}$ and so we can ignore the subdominant $m$ and $\tilde{\mu}$ terms that are $\ll M_{p}$. Therefore,
\begin{eqnarray}
{n_L}_\textrm{end}&=& Q_L \varphi^2_\textrm{end} \dot\theta_\textrm{end} \cos^2 \alpha  \nonumber \\
&\simeq& - \mathcal{O}(1) Q_L \varphi^2_\textrm{end} \frac{M_p U_{,\theta}}{f(\chi_{\rm end })\sqrt{3U_{\rm end}}} \cos^2 \alpha  \nonumber \\
&\simeq& - \mathcal{O}(1) Q_L \tilde \lambda_5 \varphi^3_\textrm{end} \sin \theta_{\rm end} /\sqrt{3 \lambda}   ~.
\end{eqnarray}
where the $\mathcal{O}(1)$ factor accounts for the approximation of the slow roll relation at the end of inflation, used in the second step; see Supp.~IB. In the last step, we assume the quartic term dominates the inflationary potential and the $\tilde \lambda_5$ coupling dominates the breaking terms. 

%For a typical parameter set that fits the inflationary constraints $\xi= 300, \lambda= 4.5 \cdot 10^{-5}$. With $\chi_0= 6 M_p, \theta_0 =0.1$ ($\varphi_{\rm end} \simeq 0.05 M_p $),  we find $n_L= 3.4 \cdot 10^{-11} M_p^3$ from the above analytic formula compared to $n_L= 1.1\cdot 10^{-10} M_p^3$ from the full numerical result. 
Numerically, we find that the $\mathcal{O}(1)$ factor is $\sim 3$, mainly originating from extending the slow roll approximation to the end of inflation, which we have defined as when the slow roll parameter is $\epsilon=1$. After inflation, the lepton number density is just red-shifted by $a^3$ with the inclusion of another $\mathcal{O}(1)$ factor $\Omega$, see Supp. IC.
\vspace{2pt}

{\bf Baryon Asymmetry Parameter}--- %
 After reheating, the generated non-zero $ n_L $  will be present in the form of neutrinos, which will be redistributed by equilibrium electroweak sphalerons, with the ratio $ n_B\simeq -\frac{28}{79} n_L $ \cite{Klinkhamer:1984di,Kuzmin:1985mm,Trodden:1998ym,Sugamoto:1982cn}. 
 To calculate the baryon to entropy ratio, we need the reheating temperature. The reheating process of Higgs inflation was first analysed in Ref. \cite{Garcia-Bellido:2008ycs, Bezrukov:2008ut}, and it has been found that the parametric resonance production of $W/Z$ plays an important role for the preheating process. It has since been determined that the preheating process is more violent than previously expected~\cite{Ema:2016dny, DeCross:2015uza, DeCross:2016cbs, DeCross:2016fdz} with unitarity being violated for  $\xi>350$. However, for such large $\xi$, the model can be UV completed in Higgs-$R^2$ inflation~\cite{Giudice:2010ka,Gorbunov:2018llf, Ema:2017rqn, Ema:2019fdd, Ema:2020zvg} for which the preheating process must be recalculated ~\cite{He:2018mgb,He:2020ivk,He:2020qcb}. In our case, we choose $\xi=300$ and thus the unitary problem is absent. A recent analysis of the preheating in Higgs inflation ~\cite{Sfakianakis:2018lzf} shows that when $\xi > 100$, the reheating happens at an e-folding number of $\sim 3$ after the end of inflation, so we adopt $\Delta N= N_\textrm{reh}- N_\textrm{end} = 3$ for simplicity. The details of reheating may be different for our case due to the doublet and triplet Higgs' mixing, and a comprehensive analysis is left for future work.
%The reheating temperature in this scenario is expected to be sufficiently large for this to occur due to the SM Higgs couplings, $ T_\textrm{reh}\sim 10^{13}-10^{14}$ GeV, as in Higgs inflation \cite{Bezrukov:2008ut}.  
For the typical parameters we consider, we find that  reheating occurs at $t_\textrm{reh} =223/H_0$ and the corresponding Hubble parameter $H_\textrm{reh}=0.0047 H_0$. From $ H_\textrm{reh}^2\simeq \frac{\pi^2}{90} g_* \frac{T_\textrm{reh}^4}{M_p^2} $ we obtain the reheating temperature 
$T_\textrm{reh} \approx 2.2\cdot10^{14}$ GeV. Considering the entropy density just after reheating  $s=\frac{2 \pi^2}{45} g_* T_\textrm{reh}^3 $ \cite{Husdal:2016haj},  the baryon asymmetry parameter is then,
\begin{eqnarray}
\eta = \frac{{n_B}}{s}  \bigg|_{\rm reh} = \eta_B^\textrm{obs} \left(\frac{{|n_L}_\textrm{end}|/M_p^{3}}{1.3\cdot 10^{-16} }  \right) \left( \frac{g_*}{112.75} \right)^{-\frac{1}{4}},
\label{baryon}
\end{eqnarray}
where $\eta_B^\textrm{obs} \simeq 8.5 \cdot 10^{-11}$ is the observed baryon asymmetry parameter \cite{Aghanim:2018eyx}. Eq.~(\ref{baryon}) shows that at the end of inflation, a lepton asymmetry of $1.3\cdot 10^{-16} M_p^{3}$ is necessary to generate the observed  baryon asymmetry, which corresponds to the example parameter sets $\tilde{\lambda}_5 = 7\cdot 10^{-15}$ for $\theta_0=0.1$, and $\tilde{\lambda}_5 = 10^{-10}$ for $\theta_0=6.5 \cdot 10^{-6}$ from numerical calculations. Note that in both of these cases, the typical parameters  escape the isocurvature limits~\cite{Byrnes:2006fr, Gordon:2000hv, Kaiser:2012ak} placed by CMB observations~\cite{Planck:2018jri}; see Supp.~ID.
% As shown in Supp.~IB, when $\tilde \lambda_5 \lesssim 10^{-10}$,  $\theta$ is approximately constant during inflation, and the lepton asymmetry density is proportional to $\tilde \lambda_5 \sin\theta$, for sufficiently small $\theta_0$. 

Comparing the quartic and dim-5 terms,  the cubic term becomes more relevant as $\varphi$ decreases. If the $\tilde \mu$ coupling becomes too large,  the lepton asymmetry starts to rapidly oscillate and predictability breaks down. On the other hand, a small $\tilde \mu$ term helps to avoid washout of the lepton asymmetry. From the analysis in Supp.~II, we require $|\tilde \mu| \lesssim 10^{-18} M_p$ for the initial $\theta_0=0.1$ to accommodate the observed baryon asymmetry.

%Considering the reference values, it is natural to assume $\lambda_H =\mathcal{O}(0.1)$ at high energy scale.
For the typical parameters in our model, we  assume $\lambda_H =0.1$, and  $\xi_H =\xi_\Delta= 300$ based on the argument above. For the other parameters, we  set $\lambda_{\Delta}= 4.5\cdot 10^{-5}$ to accommodate the inflation data, while there exists two options for $\lambda_{H \Delta}$. In the case of  $\lambda_{H \Delta} > 0$, we require $2\lambda_\Delta \xi_H -\lambda_{H\Delta}\xi_\Delta>0$ and $2\lambda_H \xi_\Delta -\lambda_{H\Delta} \xi_H>0$ to ensure the mixing of $h$ and $\Delta^0$ during inflation. The typical parameter value we can consider is $\lambda_{H \Delta}=10^{-5}$, giving the mixing angle $\alpha \simeq 0.02$. In the case of $\lambda_{H \Delta} < 0$, we need $|\lambda_{H \Delta}| < 2 \sqrt{\lambda_H \lambda_{\Delta}}$ to avoid the potential becoming unbounded from below. We can then choose  $\lambda_{\Delta}= 4.5\cdot 10^{-5}$, $\lambda_{H\Delta}=-0.001$, giving $\alpha \simeq 0.07$. 
%{\color{red} Note that in either case, the $\lambda_{\Delta}$ from the renormalization of $\lambda_{H\Delta}$ is small. However, the second case might be more stable to against the renomalization group running since $\lambda_{H \Delta}$ could also get correction for the gauge boson loops.}

Numerically, the observed baryon asymmetry is obtained  for $\lambda_5 \sim \lambda_5^{\prime}= 8.8\cdot 10^{-11}(7.9\cdot 10^{-12}) $ with $\theta_0\sim 0.1$ and the typical parameter choices given above, in the case of  $\lambda_{H\Delta}>0$ ($\lambda_{H\Delta}<0$). In addition, we obtain an upper limit $|\mu|\lesssim 15 ~(1.4)$ TeV.  Note, all these parameters are defined at the renormaliztion scale near $M_p$. Given that the $U(1)_L$ breaking term couplings are small, it is natural to consider that they originate from a spurion field which carries $U(1)_L$ charge $+2$. The $U(1)_L$ breaking terms are then generated by requiring that the spurion field obtains a VEV of order $\mathcal{O}(10^{4})$ TeV. 
\vspace{2pt}

{\bf Parameter Constraints}---  Since we expect a large reheating temperature, the triplet will be thermalized at the end of reheating, and we must consider possible washout processes. Firstly, we require that the processes $LL\leftrightarrow  HH$ is not effective,
\begin{eqnarray}
\Gamma=n\langle\sigma v\rangle\approx y^2 \mu^2/m_\Delta<H|_{T=m_\Delta}~,
\end{eqnarray}
where $ H=\sqrt{\frac{\pi^2 g_*}{90}}\frac{T^2}{M_p}$.

The triplet Higgs  generates the neutrino masses, $m_\nu\simeq y\frac{|\mu| v^2}{2m_\Delta^2}$,
where $m_\nu$ should be at least the order of the $\sim 0.05$ eV. Combining this with the above relation, we obtain $m_\Delta <10^{12}$ GeV for $m_\nu=0.05$ eV.

The other processes that are necessary to consider are $LL\leftrightarrow\Delta$ and $ HH\leftrightarrow\Delta $. They must not co-exist, otherwise the lepton number will be rapidly washed out.
However, to maintain the lepton asymmetry, the process $LL\leftrightarrow\Delta$ must be efficient while  $HH\leftrightarrow\Delta $ is out of equilibrium. This leads to the following requirement,
\begin{eqnarray}
\Gamma_{ID}(HH\leftrightarrow \Delta)|_{T=m_\Delta}<H|_{T=m_\Delta}~.
\label{con1}
\end{eqnarray}
Note that $\Gamma_{ID}(HH\leftrightarrow \Delta)|_{T=m_\Delta} \approx \Gamma_{D}(\Delta \rightarrow HH) \simeq \frac{\mu^2}{32\pi m_\Delta}$ and $v_\Delta \simeq - \frac{\mu v_{\textrm{EW}}^2}{ 2 m^2_\Delta}$. From Eq.~(\ref{con1}) one can easily get 
\begin{eqnarray}
v_\Delta \lesssim 10^{-5}~{\rm GeV} \left ( m_\Delta/{\rm TeV} \right)^{-1/2} ~,
\label{vev}
\end{eqnarray}
hence, for $m_\Delta \gtrsim 1$ TeV, we generally require that $v_\Delta \lesssim 10$ keV to prevent the washout of the lepton asymmetry.

%In Figure \ref{washout}, we show the possible parameter space where the generated lepton asymmetry is safe from washout processes. The white region subtends the allowed parameter space. 
 In Figure \ref{washout}, we depict the region of parameter space for which the generated lepton number density leads to successful Baryogenesis. The black region is excluded by requiring perturbative neutrino Yukawa $y$ couplings up to the Planck scale $M_p$ ($y \lesssim 1$). However, a small Yukawa coupling $y<0.1$ is preferred for the size of quartic coupling we consider, $\lambda_\Delta\simeq 4.5\cdot 10^{-5}$, to avoid fine-tuning at the high energy scale. The grey region is excluded by requiring that the $\mu$-term does not destroy the generated lepton asymmetry. The blue region describes the parameters that lead to washout of the lepton asymmetry. From Eq. (\ref{vev}), the blue region implies an upper limit on $v_\Delta$, namely $v_\Delta \lesssim 10$ keV.

 %%%%%%%%%%%%%%%%%%%%%%%%%%%%%%%%%%%%%%%%%%%%%%%%%%%%%%
 %%%%%%%%%%%%%%%%%%%%%%%%%%%%%%%%%%%%%%%%%%%%%%%%%%%%%%
 \begin{center}
 \includegraphics[width=0.8\columnwidth]{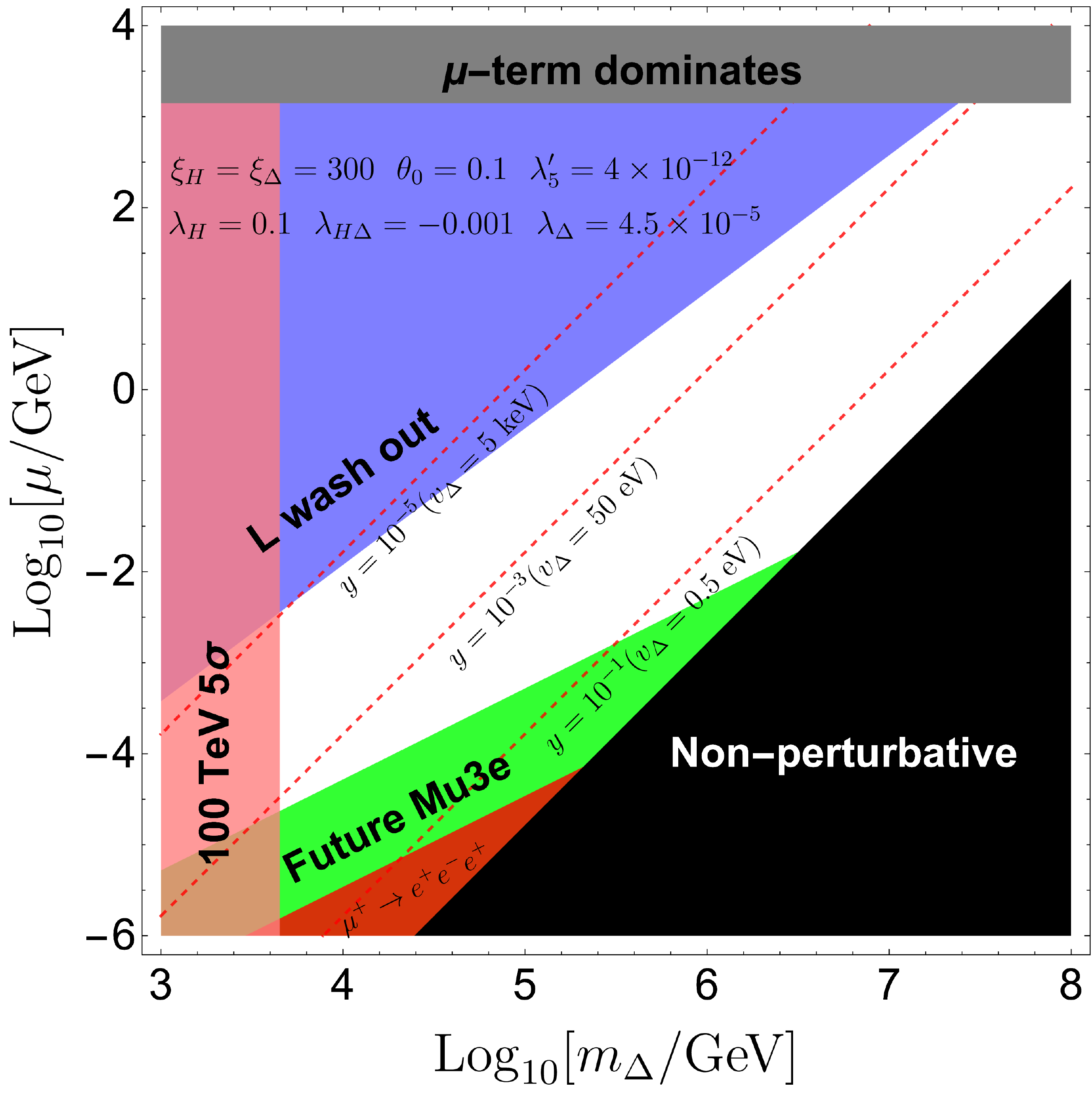}
 \captionof{figure}{The allowed region of parameter space is depicted (White), avoiding washout processes (Blue), cubic term domination of $U(1)_L$ breaking (Grey), and non-perturbative neutrino Yukawa couplings (Black). The red region denotes the current limits from lepton violating decays \cite{SINDRUM:1987nra}, with green indicating the future Mu3e experimental sensitivity \cite{Perrevoort:2018ttp}. The future 100 TeV collider constraints are depicted by the light red region \cite{Du:2018eaw}.}
 \label{washout} 
 \end{center}
 %%%%%%%%%%%%%%%%%%%%%%%%%%%%%%%%%%%%%%%%%%%%%%%%%%%%%%
 %This includes constraints from non-perturbativity of the neutrino Yukawa couplings (Black)
 %%%%%%%%%%%%%%%%%%%%%%%%%%%%%%%%%%%%%%%%%%%%%%%%%%%%%% 

There is an additional limit from  precision measurements on the vacuum expectation value $\langle\Delta^0\rangle$, namely, it must be less than a few GeV. 
%In our model, the triplet Higgs vacuum expectation value is generically small {\color{red}in order to explain the neutrino masses}. 
LHC searches apply lower bounds on the masses of the triplet Higgs components, for example the current limit on the mass of the doubly charged Higgs is $ \sim800 $ GeV~\cite{ATLAS:2017xqs}, and thus we only depict triplet masses $\geq 1$ TeV. This limit may be increased at the upgraded high luminosity LHC or at future colliders \cite{Du:2018eaw}. 
%We have found that stable Q-balls are unlikely to form \cite{Coleman:1985ki, Lee:1991ax,Kusenko:1997zq, Kusenko:1997ad}, which is discussed in Supp.~IF. There may be gravitational wave implications~\cite{White:2021hwi}.
Assuming the Yukawa couplings are of the same order of magnitude, the lepton flavor violating processes induced by the doubly-charged Higgs already provide a limit on the parameter space  \cite{SINDRUM:1987nra, Han:2021nod}. The current limit will be improved by two orders of magnitude by the upcoming Mu3e experiment \cite{Perrevoort:2018ttp}.
\vspace{2pt}

{\bf Concluding Remarks}--- 
We have shown that the introduction of the triplet Higgs of the Type II seesaw mechanism to the SM provides a simple framework in which  inflation, neutrino masses, and the baryon asymmetry are all explained. We now summarize the unique combination of phenomenological predictions of this model:
\begin{enumerate}
\item  %In contrast to the parameter space usually considered in the Type II seesaw, to explain the baryon asymmetry we allow only a small part of usual the parameter space. The vacuum value $v_\Delta$ can only be accommodated within the range 10 keV - eV, and subsequently the decay channel of the triplets is predominantly leptonic. This fact provides valuable insight for triplet Higgs searches at future collider. 
Depending upon the vacuum value of the triplet and its Yukawa couplings, the triplet Higgs can decay mainly into  gauge bosons or leptons. In our model, $v_\Delta$  can only be accommodated within the range 10 keV - eV, with the upper limit ensuring  lepton asymmetry washout effects are negligible. Importantly, for this $v_\Delta$ range the triplet Higgs dominantly decays into leptons. If  we observed the  triplet Higgs in such a channel, it would provide a smoking gun for our model.

\item The associated doubly-charged Higgs directly leads to lepton flavor violating processes such as $\mu \rightarrow e \gamma$, and $\mu \rightarrow e e e$. The current experimental limits already provide constraints on the triplet Higgs properties, see Fig. 1, with future experiments such as  Mu3e to improve upon the $\mu \rightarrow e e e$ limits by two orders of magnitude. Thus, the allowed parameter space of our model will be tested in near future.

\item In this model, the observed neutrino masses are of the Majorana type. This is in contrast to  models which include right-handed neutrinos, where the observed neutrinos can have both Dirac and Majorana type mass terms. Thus, our model can be probed at near future neutrinoless double beta decay experiments. In addition, the baryon asymmetry generated in this model is independent of the leptonic $\mathcal{CP}$ phase, with $\mathcal{CP}$ spontaneously broken at early times of the universe. There is currently conflicting measurements of the leptonic $\mathcal{CP}$ phase coming from the T2K and NOvA experiments, with T2K disfavouring a $\mathcal{CP}$ conserving angle \cite{T2K:2019bcf}, which is inconsistent with the NOvA result \cite{NOvA:2021nfi}. In this context, our model provides an interesting theoretical possibility for Leptogenesis. 
%Thus, a vanishing the observation of a vanishing $\mathcal{CP}$ phase in the neutrino sector still allows successful Leptogenesis. Although is not a must...
%On the neutrino physics side,  our model predicts that the neutrinos must be of the majorana type. This is in contrast to  models which include right-handed neutrinos, where the observed neutrinos can have both Dirac type or majorana type mass terms. The model presented here provides this unique prediction which will be probed at near future doubly-beta decay experiments. Given many on-going such experiments, we believe our result provides more motivation for the consideration of majorana type neutrinos. In addition, comparing with usual thermal leptogenesis,  our model can generate the correct baryon asymmetry independent of the $\mathcal{CP}$ phase in the neutrino sector. Therefore, if future neutrino experiments find that the $\mathcal{CP}$ is conserved in the neutrino sector (currently there is conflicting evidence coming from T2K, which excludes $\mathcal{CP}$ conserving angle at 90\% CL, and the Nova result which prefers a vanishing $\mathcal{CP}$ phase),  our model provides an interesting theoretical possibility for Leptogenesis. 

\item We assume that the inflationary period is induced by two scalar fields. Generally, such inflationary setups can generate non-trivial non-Gaussian features ~\cite{Kaiser:2012ak}. In addition, a sizable isocurvature  signature could be produced if considerable washout effects are allowed at late times. These possibilities may be probed by future observations.
%On the cosmological side,  our model is a two field inflation mechanism with non-minimal couplings. Such a model could provide a non-Gaussian signatures in particular parameter spaces, that do not assume the simple slow roll region characterised by the Lebedev-Lee solution. It can in principle be extended to a more general parameter space and the predicted non-Gaussian signatures could be determined. Additionally, the inflationary dynamics could produce sizable isocurvature perturbations in certain parameter regions, which may be probed by future observational experiments. XXXX
\end{enumerate}

{\bf Acknowledgment.}---%
We would like to thank Tsutomu T. Yanagida, Misao Sasaki, Shi Pi and Jiajie Ling for their helpful discussions. C. H. is supported by the Guangzhou Basic and Applied Basic Research Foundation under Grant No. 202102020885, and the Sun Yat-Sen University Science Foundation. NDB is supported by IBS under the project code, IBS-R018-D1.  The work of HM was supported by the Director, Office of Science, Office of High Energy Physics of the U.S. Department of Energy under the Contract No. DE-AC02-05CH11231, by the NSF grant PHY-1915314, by the JSPS Grant-in-Aid for Scientific Research JP20K03942, MEXT Grant-in-Aid for Transformative Research Areas (A) JP20H05850, JP20A203, by WPI, MEXT, Japan, and Hamamatsu Photonics, K.K.

%\bibliography{main} 

%\newpage
%\bibliographystyle{utcaps_mod}
\bibliography{bibly}

\clearpage
\newpage
\maketitle
\onecolumngrid
\begin{center}
\textbf{\large Affleck-Dine Leptogenesis from Higgs Inflation} \\ 
\vspace{0.05in}
{ \it \large Supplemental Material}\\ 
{By Neil D. Barrie, Chengcheng Han, Hitoshi Murayama}
\vspace{0.05in}
\end{center}
\onecolumngrid
%%%%%%%%%% Merge with Supplemental material %%%%%%%%%%
\setcounter{equation}{0}
\setcounter{figure}{0}
\setcounter{table}{0}
\setcounter{section}{0}
\setcounter{page}{1}
\makeatletter
\renewcommand{\theequation}{S\arabic{equation}}
\renewcommand{\thefigure}{S\arabic{figure}}

\section{Inflationary Observables}
\label{Inf_obs}

\subsection{Single field inflation approximation}
In this subsection, we consider the inflationary dynamics with the assumption of a single field approximation, requiring $\tilde \lambda_5/M_p \ll \lambda$ and $\tilde \mu \ll \lambda \varphi$. In transforming from the Jordan to Einstein frame, the field $ \varphi $ no longer has a canonically normalised kinetic term. To rectify this, we make the following field redefinition,
\begin{equation}
\frac{d\chi}{d\varphi}
\;=\; \dfrac{\sqrt{(6\xi^2 \varphi^2/M_p^2)+\Omega^2}}{\Omega^2}
\;,
\end{equation}
where $ \Omega^2=1+\xi\left(\dfrac{\varphi}{M_p}\right)^2 $. The field redefinition of $\chi$ in terms of $ \phi $ is given by, 
\begin{eqnarray}
\chi(\varphi) &=& 1/\sqrt{\xi} (\sqrt{1+6\xi}~ \sinh^{-1}(\sqrt{\xi+6\xi^2}  \varphi) \nonumber \\
 && -\sqrt{6\xi} ~ \sinh^{-1} (\sqrt{6\xi^2}\varphi/\sqrt{1+ \xi \varphi^2})~.
\end{eqnarray} 
There are three distinct regimes that can be seen in the evolution of the relation between $ \chi $ and $ \varphi $, namely,
 \begin{equation}
 \dfrac{\chi}{M_p}
 \approx
 \left\{
 \begin{array}{lll}
 \dfrac{\varphi}{M_p}
 & \mbox{for $\dfrac{\varphi}{M_p} \ll\dfrac{1}{\xi}$} 
 & \mbox{(radiation-like)}
 \\
 \sqrt{\dfrac{3}{2}}\,\xi\left(\dfrac{\varphi}{M_p}\right)^2 
 & \mbox{for $\dfrac{1}{\xi}\ll \dfrac{\varphi}{M_p} \ll \dfrac{1}{\sqrt{\xi}}$\quad} 
 & \mbox{(matter-like)}
 \\
 \sqrt{\dfrac{3}{2}} \ln\Omega^2 =
 \sqrt{\dfrac{3}{2}} \ln\left[1+\xi\left(\dfrac{\varphi}{M_p}\right)^2\right] \quad
 & \mbox{for $\dfrac{1}{\sqrt{\xi}}\ll \dfrac{\varphi}{M_p}$} 
 & \mbox{(inflation)}
 \end{array}
 \right.
 \label{chi-varphi-relation_sm}
 \end{equation}

Translating this to the effects on the scalar potential, we can see how the non-minimal coupling leads to a flattening of the potential in the large field limit. The scalar potential in each regime is signified by,
\begin{equation}
U(\chi) \approx
\left\{
\begin{array}{lll}
\dfrac{1}{4}\lambda\chi^4 
& \mbox{for $\dfrac{\chi}{M_p}\ll\dfrac{1}{\xi}$}
& \mbox{(radiation-like)}
\\
\dfrac{1}{2} m_{S}^2 \chi^2 
& \mbox{for $\dfrac{1}{\xi} \ll \dfrac{\chi}{M_p} \ll 1$\quad}
& \mbox{(matter-like)}
\\
\dfrac{3}{4}m_{S}^2 M_p^2
\left(1-e^{-\sqrt{\frac{2}{3}}(\chi/M_p)}\right)^2 \quad
& \mbox{for $1\ll \dfrac{\chi}{M_p}$}
& \mbox{(inflation)}
\end{array}
\right.
\label{Uchi_sm}
\end{equation}

Considering the inflationary potential parametrised by $ \chi $ above, we see that this is exactly the Starobinsky inflationary potential. This scenario has the following characteristic slow roll parameters,
\begin{eqnarray}
\epsilon\simeq \frac{3}{4N_*^2}~, ~~\textrm{and} ~~\eta\simeq \frac{1}{N_*}~,
\label{slowroll_sm}
\end{eqnarray} 
where $ N_* $ is the number of e-foldings of expansion from the horizon crossing to the end of the inflation. Thus, the spectral index and tensor to scalar ratio are given by
\begin{eqnarray}
n_s\simeq 1-\frac{2}{N_*}~, ~~\textrm{and} ~~r\simeq \frac{12}{N_*^2}~,
\label{inflation_obs_sm}
\end{eqnarray} 
respectively.
%The value of $ N_* $ is related to the field displacement of the inflaton, and approximately given by,
%\begin{eqnarray}
%N\approx -\int_{\chi_0}^{\chi_f} d\chi \frac{U}{U_{,\chi}}~\approx \frac{3}{4} e^{\sqrt{\frac{2}{3}}(\chi_0/M_p)},
%\label{efolds_sm}
%\end{eqnarray} 
%where $ \chi_0 $ and $ \chi_f $ are the initial and final inflaton field values respectively. For $ N>50 $, it is required that $ \chi_0 \gtrsim 5.14 M_p $.

      \begin{figure*}[h]
      	\centering
      	\includegraphics[width=0.7\textwidth]{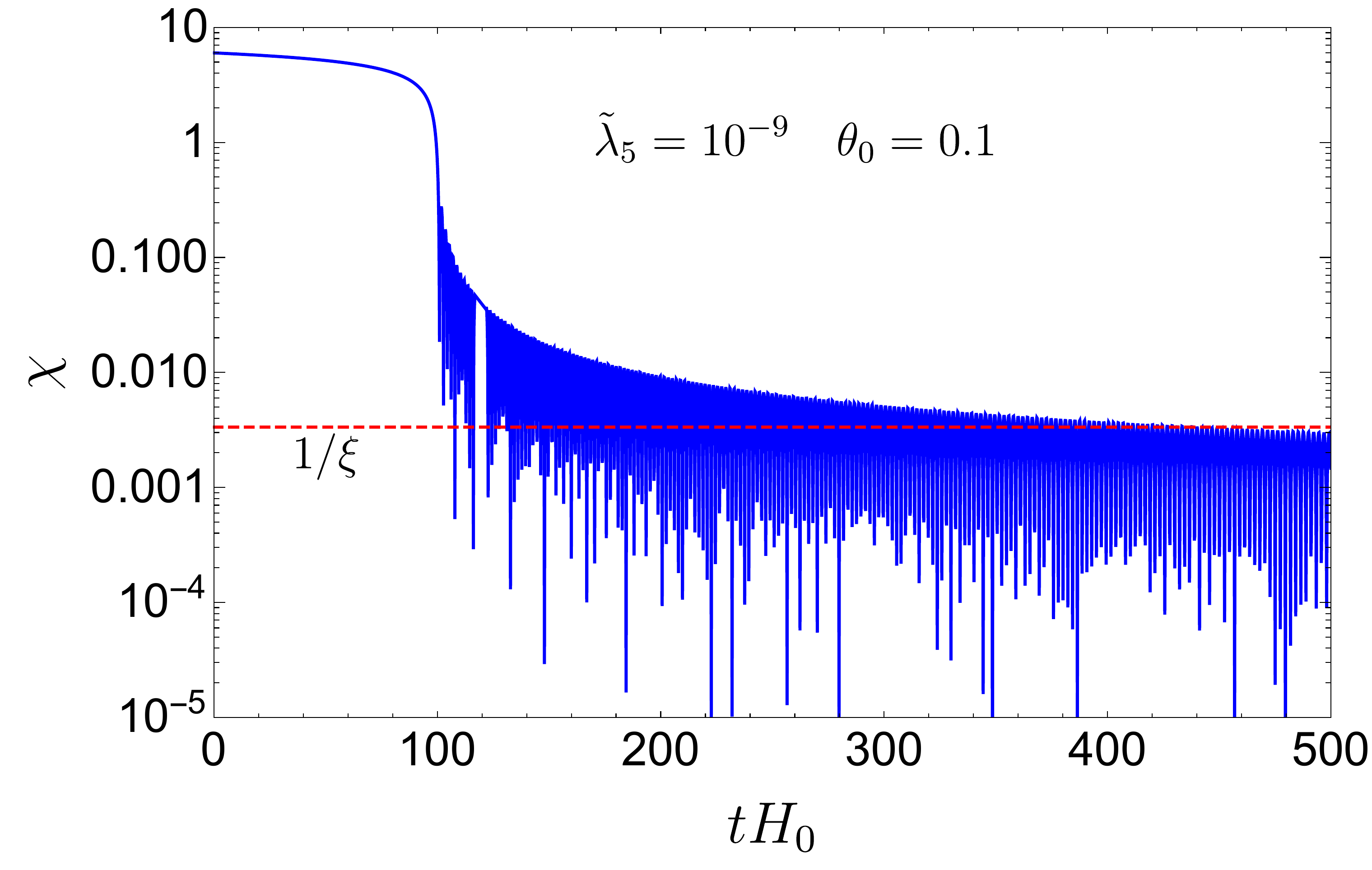}
      	\caption{The evolution of $\chi$ during and after inflation, in Planck units. The input parameters are fixed to $ \xi= 300$ and $\lambda=4.5 \cdot 10^{-5}$, with initial conditions  $\chi_0=6.0 M_p$, $\dot \chi_0=0$,  and $\dot \theta_0=0 $ chosen. We require $\tilde\mu$ to be sufficiently small to not affect the dynamics, and $H_0$ is defined as $H_0\equiv m_S/2$. }
      	\label{chi}
      \end{figure*}

We can now compare the inflationary predictions to the current Planck results,
\begin{eqnarray}
\label{ns}
n_s = 0.9649\pm 0042~~{\rm (68\% C.L.)}~,\\
\label{r}
r_{0.002} < 0.056~~{\rm (95\% C.L.)}~,
\end{eqnarray}
for the $\Lambda-$CDM$+r$ model. Comparing this with the Starobinsky inflation predictions, we see that they are extremely well fitted for the required number of inflationary e-folds $ 50<N_*<60 $ .
 
In addition, agreement with the scalar perturbations observed in the CMB places a requirement on the Starobinsky mass scale $ m_S $. The following parameter relation is necessitated, 
\begin{eqnarray}
\frac{\lambda}{\xi^2}\simeq 5\cdot 10^{-10} ~,
\end{eqnarray} 
leading to the Starobinsky mass scale $ m_S\simeq 3 \cdot 10^{13} $ GeV.

As an illustration, we provide a plot for the evolution of the $\chi$ field during and after inflation. The input parameters are fixed to $ \xi= 300$ and  $\lambda=4.5 \cdot 10^{-5}$. We take $\tilde\mu$ to be sufficiently small to not affect the dynamics, and the following initial conditions are chosen, $\chi_0=6.0 M_p$, $\dot \chi_0=0$,  and $\dot \theta_0=0 $ with $H_0\equiv m_S/2$. In Fig.~\ref{chi}, we include the line  $\chi =  M_p/\xi$, which indicates the end of the matter-like evolution. The plot shows that the end of inflation occurs near $tH_0=100$ and therefore the horizon crossing happens at $tH_0 \approx 40-50$.  After inflation, the universe quickly enters into the matter-like epoch {until} $t H_0 \approx 400$.

\subsection{Dynamics of $ \theta $}
\label{theta_dyn}

Consider the equation of motion for  $ \theta $,
\begin{eqnarray}
 \ddot{\theta} + \frac{f'(\chi)}{f(\chi)}  \dot \theta \dot \chi + 3 H \dot \theta +\frac{1}{f(\chi)} U_{,\theta} =0 ~.
\label{thetaEOM}
\end{eqnarray} 
At the beginning of inflation, $\dot \theta$ increases until it reaches the slow roll regime. Then we obtain,
\begin{eqnarray}
    \dot{\theta} \simeq -\frac{U_{,\theta}}{f(\chi)\sqrt{3U}}  ~.
    \label{dottheta_inf11}
\label{thetaEOM2}
\end{eqnarray}

  \begin{figure}[t!]
      \begin{subfigure}
      	\centering
      	\includegraphics[width=0.48\textwidth]{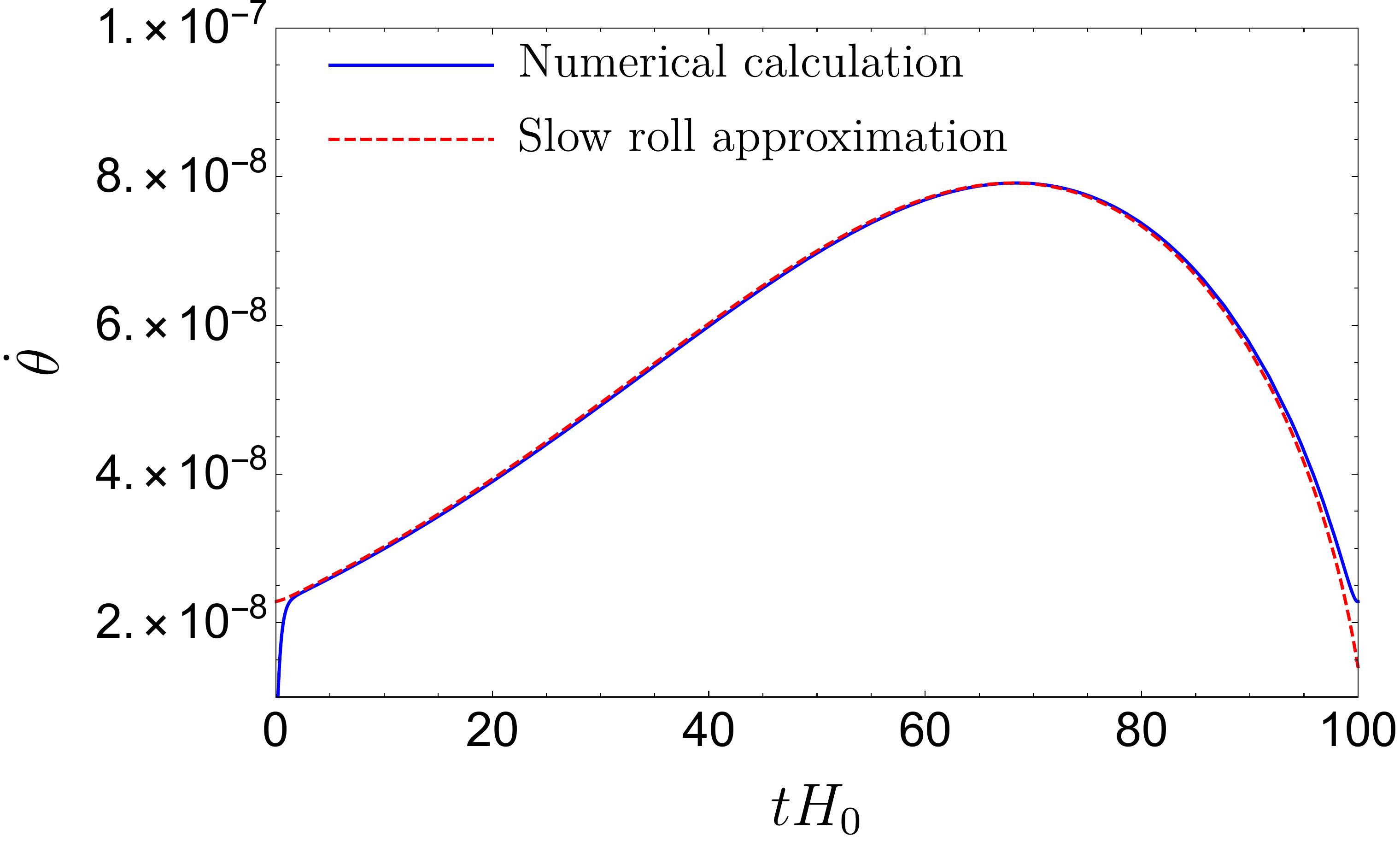}
      	\end{subfigure}
      	\hfill 
      	   \begin{subfigure}
      	   	\centering
      	   	\includegraphics[width=0.45\textwidth]{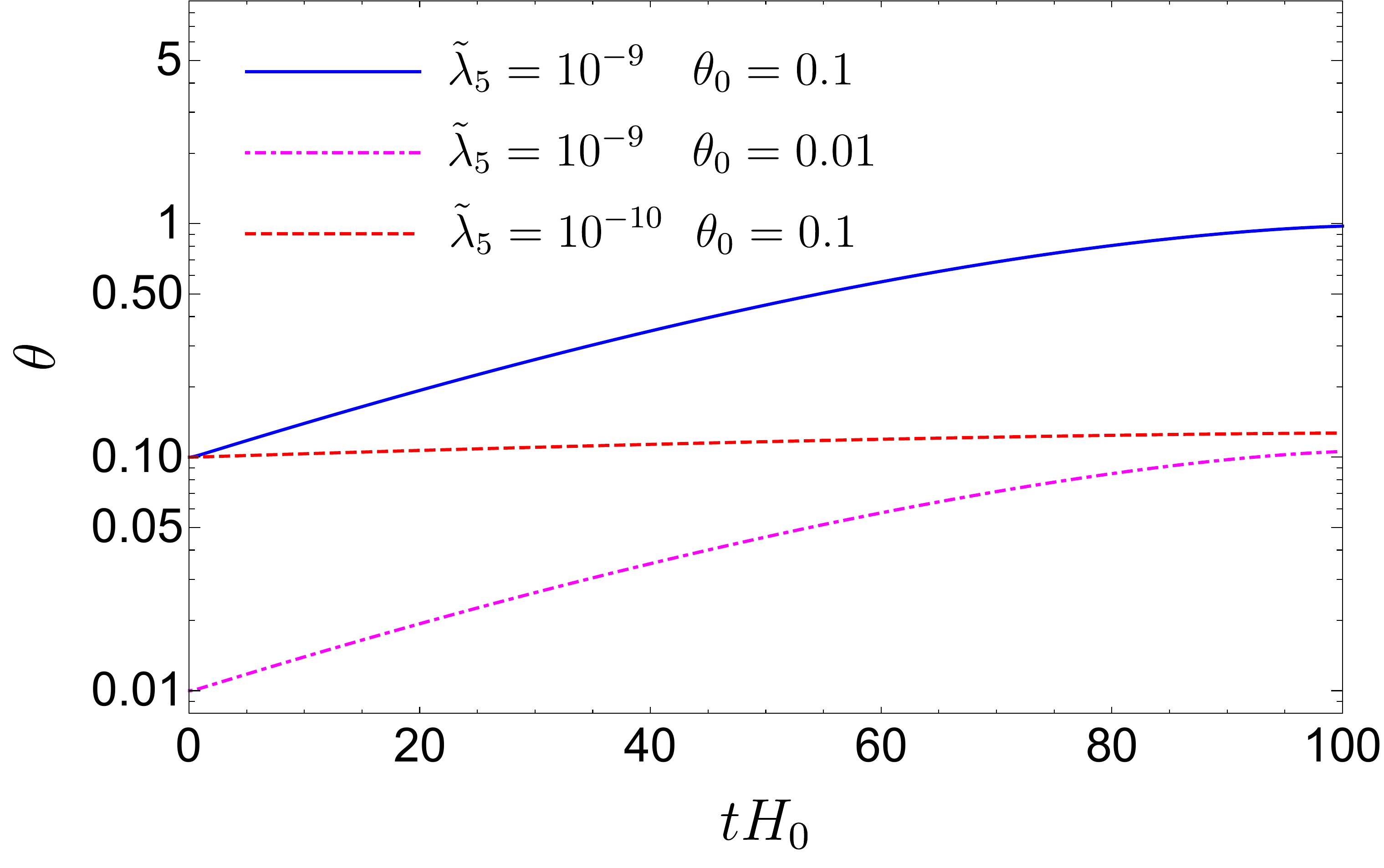}
      	   	\end{subfigure}
      	   
      	\caption{  The dynamics of (left) $\dot \theta$ and (right) $\theta$ during inflation are depicted, in Planck unit. The left figure includes a comparison of the exact numerical result and analytical result given in Eq.~(\ref{thetaEOM2}) with $\tilde \lambda_5= 10^{-9}$. The other input parameters are fixed to $ \xi= 300$ and $\lambda=4.5 \cdot 10^{-5}$, with initial conditions  $\chi_0=6.0 M_p$, $\dot \chi_0=0$,  and $\dot \theta_0=0 $ chosen. We choose $\tilde\mu$ to be sufficiently small to not affect the dynamics. }
      	\label{theta}
      \end{figure}

In Fig.~\ref{theta}, we show the evolution of $\theta$ and $\dot \theta$ during inflation, including the estimation of  $\dot \theta$ presented in Eq.~(\ref{thetaEOM2}). We find that $\dot \theta$ is well described by the slow roll approximation. Due to the non-vanishing $\dot \theta$, $\theta$ increases during inflation. However, as shown in the right panel of the Fig.~\ref{theta}, for fixed $\theta_0$, the rate of increase is proportional to $\tilde \lambda_5$, which is indicated by Eq.~(\ref{thetaEOM2}).

\subsection{ Lepton Number Density After Inflation} % 
%Now that we have detailed the inflationary dynamics, 
Consider the calculation of the lepton number density,
\begin{eqnarray}
n_L=Q_L \varphi^2 \dot\theta \cos^2 \alpha    ~.
\end{eqnarray}
It is clear that we must determine the dynamics of $ \dot{\theta} $. We assume that the initial $ \dot{\theta}_0 $ is zero, with a non-zero initial $ \theta=\theta_0 $. It will become evident that the sign of the resultant asymmetry is dependent upon the choice of $ \theta_0 $. 

Considering Eq. (\ref{thetaEOM}), we see that $ \dot\theta $ will quickly enter the slow roll regime after a few e-folds, giving,
   \begin{eqnarray}
    \dot{\theta} \simeq -\frac{U_{,\theta}}{f(\chi)\sqrt{3U}} ~
    \label{dottheta_inf11}
    \end{eqnarray}
in the slow roll approximation.

Once inflation comes to an end ($\chi_e \simeq 0.67 M_p $), the oscillatory epoch begins and the universe behaves as approximately matter-like. The inflaton potential during this stage is shown in Supp.~IA. We define the lepton asymmetry at the end of the inflationary epoch as,
\begin{eqnarray}
{n_L}_\textrm{end}= Q_L \varphi^2_\textrm{end} \dot\theta_\textrm{end} \cos^2 \alpha    ~.
\end{eqnarray}

From Eq. (\ref{thetaEOM}) we find that, 
\begin{eqnarray}
\frac{d [  a^3 f(\chi) \dot \theta]}{dt} = \frac{d [ a^3 n_L/(Q_L\Omega^2)]}{dt}  = a^3 U_{,\theta}
\label{dlepton}
\end{eqnarray}
Consider the dynamics of this relation after inflation. During the oscillation phase, if the potential $ U_{,\theta}$ red-shifts faster than matter-like $a^{-3}$, then we can safely ignore it and $a^3 n_L/(Q_L\Omega^2)$ will be conserved. This is guaranteed when the $U(1)_L$ breaking term in $V(\varphi,\theta)$ is a polynomial function of $\varphi$ larger than four because the quartic term of $\varphi$ in $V(\varphi,\theta)$ is equivalent to the mass term of the $\chi$ field, which red-shifts as $a^{3}$ after inflation. Thus, we consider the $\tilde{\lambda}_5$ to be the dominant $U_L$ breaking term at large field values. If instead the cubic term dominates,  the $U_L$ breaking term becomes more relevant with the expansion of the universe and thus destroys the lepton number generated during inflation. Later we will show that a small cubic term is required to avoid lepton asymmetry wash-out effects after reheating.

\begin{figure*}[h]
      	\centering
      	\includegraphics[width=0.7\textwidth]{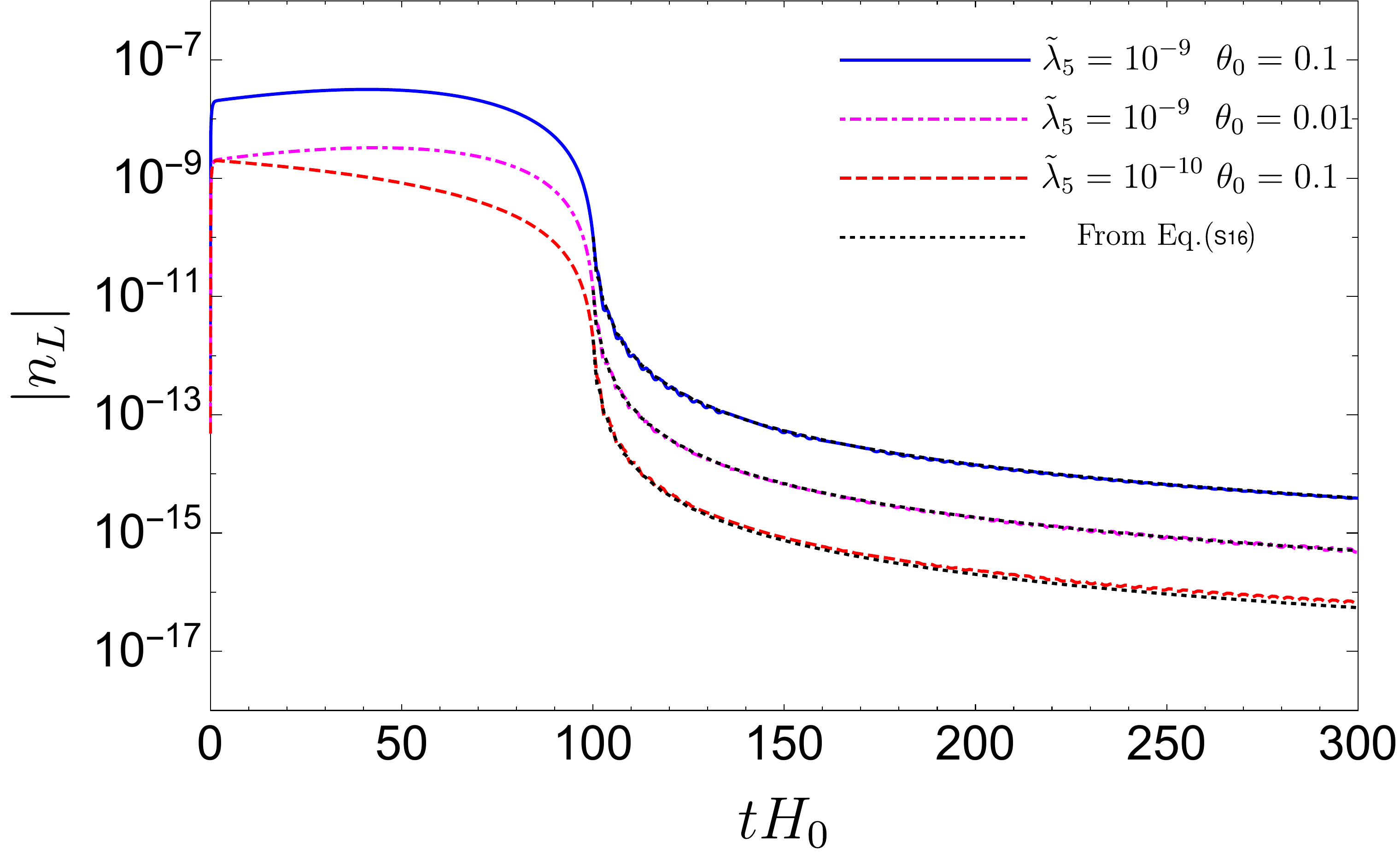}
      	\caption{Comparison of the $ n_L $ derived from full numerical calculations (Coloured lines) and the analytical estimation (Black dotted lines) in Eq.~(\ref{lepton}). For varying $\theta_0$ and $\tilde{\lambda}_5$, in Planck units. The fixed input parameters are $ \xi= 300$, $\lambda=4.5 \cdot 10^{-5}$. We take $\tilde\mu$ to be sufficiently small to not affect the dynamics, and the following initial conditions are chosen, $\chi_0=6.0 M_p$, $\dot \chi_0=0$,  and $\dot \theta_0=0$. }
      	\label{numerical}
      \end{figure*}

The relation in Eq.~(\ref{dlepton}) shows that after inflation the lepton number density is red-shifted by the usual scale factor  dependence alongside an $\mathcal{O}(1)$ factor $1/\Omega^2$. Thus, the lepton asymmetry at any time after inflation can be estimated by,
\begin{eqnarray}
n_L(t) = {n_L}_\textrm{end} \frac{\Omega^2(\chi)}{\Omega^2(\chi_\textrm{end})} \left (\frac{a}{a_\textrm{end}}\right)^{-3}.
\label{lepton}
\end{eqnarray}
The accuracy of this relation is demonstrated in Fig.~\ref{numerical}, which compares the numerical simulations to the analytical result above. The input parameters have been chosen to fit the inflationary observables.

In the above analysis, we assume that the mixing angle $\alpha$ is fixed throughout the inflationary and 
the oscillation epochs.   In fact after inflation, the $\Omega^2$ factor quickly approaches 1 and the potential can be approximated by $V(h,\Delta^0)$. The direction of the minimum of this potential has angle $\beta$, satisfying,
\begin{eqnarray}
\frac{ \rho_{H}}{\rho_{\Delta}} \equiv \tan \beta = \sqrt{\frac{2\lambda_\Delta -\lambda_{H\Delta} }{ 2\lambda_H -\lambda_{H\Delta}  }}~,
\end{eqnarray}
for $2\lambda_\Delta -\lambda_{H\Delta}>0$ and $2\lambda_H -\lambda_{H\Delta}>0 $. Interestingly, for $\xi_H = \xi_\Delta$, the two mixing angles $\alpha$ and $\beta$ converge and we can  utilise the same mixing angle for the inflationary  and oscillation stage. If $\xi_H \neq \xi_\Delta$, the $\alpha$ and $\beta$ are generally different and the oscillation stage after inflation requires a dedicated analysis. However, we believe that the lepton asymmetry is only negligibly affected in this case, since almost all of lepton asymmetry is generated during inflation, which is subsequently red-shifted by the expansion of the universe. The detail of the oscillation stage might affect the preheating and the reheating temperature. Such analysis is beyond the scope of this paper, and as such we adopt  $\xi_H = \xi_\Delta$ for simplicity.

\subsection{ Isocurvature Fluctuations}
Since our model contains multiple scalar fields, we must consider the observational limits from isocurvature perturbations. In doing this calculation we have followed the works of [17] and [20], and the formalism used in [85].

In our model, besides the $\chi$ field, the only relevant dynamics are from the $\theta$ field which generates the baryon asymmetry during the inflation. To calculate the isocurvature perturbations, we need to consider the dynamics of two-field inflation $(\chi,\theta)$. A general action in the Einstein frame can be written as,
\begin{eqnarray}
S_{\rm E} = \int d^4 x \sqrt{-g} \left[\frac{M_p^2}{2}R- \frac{1}{2} h_{IJ} g^{\mu\nu} \partial_\mu \phi^I \partial_\nu \phi^J- V(\phi) \right]  ~,
\end{eqnarray}
where $h_{IJ}$ is the metric in field space.
We define $\phi^I(x^\mu)= \varphi^I(t) + \delta \phi^I(x^\mu)$, which leads to the equation of motion,
\begin{eqnarray}
\mathcal D_t \dot \varphi^I + 3 H \dot \varphi^I + h^{IJ} V_{,K} =0  ~,
\end{eqnarray}
where $\mathcal D_t$ is the covariant directional derivative 
$\mathcal D_t A^I\equiv \dot \varphi^I \mathcal D_J A^I = \dot A^I + \Gamma^I_{JK} A^J\dot \varphi^K $.
The Mukhanov-Sasaki variables are defined as,
\begin{eqnarray}
Q^I= \delta \phi^I +\frac{\dot\varphi^I}{H} \psi ~.
\end{eqnarray}
Then we have,
\begin{eqnarray}
\mathcal D_t^2 Q^I + 3 H \mathcal D_t Q^I + \left[ \frac{k^2}{a^2} \delta^I_J + \mathcal M^I_J -\frac{1}{M_p^2 a^3} \mathcal D_t \left(  \frac{a^3}{H}  \dot \varphi^I \dot \varphi_J \right) \right] Q^J =0  ~,
\end{eqnarray}
where 
\begin{eqnarray}
\mathcal M^I_J \equiv h^{I K} (\mathcal D_J \mathcal D_K V) - \mathcal R^I_{LMJ} \dot \varphi^L \dot \varphi^M  ~.
\end{eqnarray}
The adiabatic field $\sigma$ and its direction  $\hat \sigma^I$ can be calculated as following,
\begin{eqnarray}
&& \dot \sigma^2 = h_{IJ} \dot \varphi^I \dot \varphi^J  ~, \\
&& \hat \sigma^I \equiv \frac{\dot \varphi^I}{\dot \sigma}~.
\end{eqnarray}
Now we have
\begin{eqnarray}
&& H^2 =\frac{1}{3 M_p^2} \left[  \frac{1}{2} \dot \sigma^2 + V \right]  ~, \\
&& \dot H = -\frac{1}{2 M_p^2} (\dot \sigma)^2  ~,\\
&&\ddot \sigma + 3 H \dot \sigma + V_{,\sigma} =0 ~,
\end{eqnarray}
where $V_{,\sigma} \equiv \hat \sigma^I V_{,I}$.
The entropy direction can then be defined as,
\begin{eqnarray}
&& \omega^I \equiv \mathcal D_t \hat \sigma^I  ~, \\
&& \hat s^I \equiv \frac{\omega^I}{\omega} ~,
\end{eqnarray}
where $\omega= \sqrt{h_{I J}\omega^I \omega^J}$. \\
The corresponding slow roll parameters are,
\begin{eqnarray}
\epsilon & \equiv& -\frac{\dot H}{H^2} = \frac{3 \dot \sigma^2}{\dot \sigma^2+2V} ~, \\
\eta_{\sigma\sigma} &\equiv& M_p^2 \frac{\hat \sigma_I \hat \sigma^J \mathcal M^I_J}{V} ~, \\
\eta_{ss} &\equiv& M_p^2 \frac{\hat s_I \hat s^J \mathcal M^I_J}{V} ~. 
\end{eqnarray}
The adiabatic and isocurvature perturbations are parameterized as,
\begin{eqnarray}
\mathcal{R}_c &=& (H/\dot \sigma) \hat \sigma_I Q^I ~, \\
\mathcal{S} &=& (H/\dot \sigma) \hat s_I Q^I ~. \\
\end{eqnarray}
After horizon crossing, $\mathcal{R}_c$ and $\mathcal{S}$ evolve as follows,
\begin{eqnarray}
\dot{\mathcal{R}}_c &=& \alpha H \mathcal S + \mathcal O (\frac{k^2}{a^2 H^2})  ~, \\
\dot{\mathcal{S}} &=& \beta H \mathcal{S} +\mathcal{O} (\frac{k^2}{a^2 H^2})  ~, 
\end{eqnarray}
where $\alpha =\frac{2\omega(t)}{H(t)}$ and $\beta(t) =-2 \epsilon -\eta_{ss}+\eta_{\sigma\sigma} -\frac{4}{3} \frac{\omega^2}{H^2}$.

Then one can define the transfer functions as
\begin{eqnarray}
\left(
\begin{array}{c}
    \mathcal{R}_c   \\
    \mathcal{S} 
\end{array}
\right)
=
\left(
\begin{array}{cc}
    1 & T_{\mathcal RS}  \\
    0 & T_{\mathcal SS} 
\end{array}
\right)
\left(
\begin{array}{c}
    \mathcal{R}_c   \\
    \mathcal{S} 
\end{array}
\right)_* ~,
\end{eqnarray}
with
\begin{eqnarray}
T_\mathcal{RS}(t_*, t) &=&\int^t_{t_*} dt^\prime 2 \omega(t^\prime) T_\mathcal{SS}(t_*, t)  ~,\\
T_\mathcal{SS}(t_*, t) &=&\exp\left[ \int^t_{t_*} dt^\prime \beta(t^\prime) H(t^\prime)   \right]  ~.
\end{eqnarray}
The correlation of the curvature and isocurvature modes is typically defined as, 
\begin{eqnarray}
\cos \Delta &\equiv& {T_\mathcal{RS}}/(1+ T^2_\mathcal{RS})^{1/2} ~.
\end{eqnarray}
From the Planck data,  a limit can be placed on this parameter,  $\cos \Delta \lesssim 0.1 [86]$. 

For our model, we have,
\begin{eqnarray}
\phi^1 =\chi,~~~ \phi^2=\theta, ~~~ h_{IJ} =h_{IJ}(\chi) = 
\left(
\begin{array}{cc}
    1 & 0  \\
    0 &  f(\chi) 
\end{array}
\right)  ~,
\end{eqnarray}
where {$f(\chi) \equiv  \frac{\varphi(\chi)^2 \cos^2 \alpha}{\Omega^2(\chi)}$}.

Now we have,
\begin{eqnarray}
\Gamma^1_{ij} = 
\left(
\begin{array}{ccc}
 0 & 0    \\
 0 & -\frac{1}{2} f^\prime     
\end{array}
\right),~~~
\Gamma^2_{ij} = 
\left(
\begin{array}{ccc}
 0 & \frac{f^\prime}{2 f}      \\
 \frac{f^\prime}{2 f}  & 0     
\end{array}
\right) ~,
\end{eqnarray}
The only non-vanishing components of the Riemann tensor $R^l_{kji}$ are,
\begin{eqnarray}
R^1_{212} =-R^1_{221} =-R^2_{112} =R^2_{121} = \frac{{f^\prime}^2}{4 f} -\frac{f^{\prime\prime}}{2}   ~,
\end{eqnarray}
subsequently the Ricci curvature tensor $R_{ji}$ and Ricci scalar $R$ are,
\begin{eqnarray}
R_{12} &=&R_{21} = \frac{{f^\prime}^2- 2 f f^{\prime\prime} }{4f^2} ~, \\
R &=& \frac{{f^\prime}^2- 2 f f^{\prime\prime} }{2f^2}  ~.
\end{eqnarray}

\begin{figure*}[h]
      	\centering
      	\includegraphics[width=0.55\textwidth]{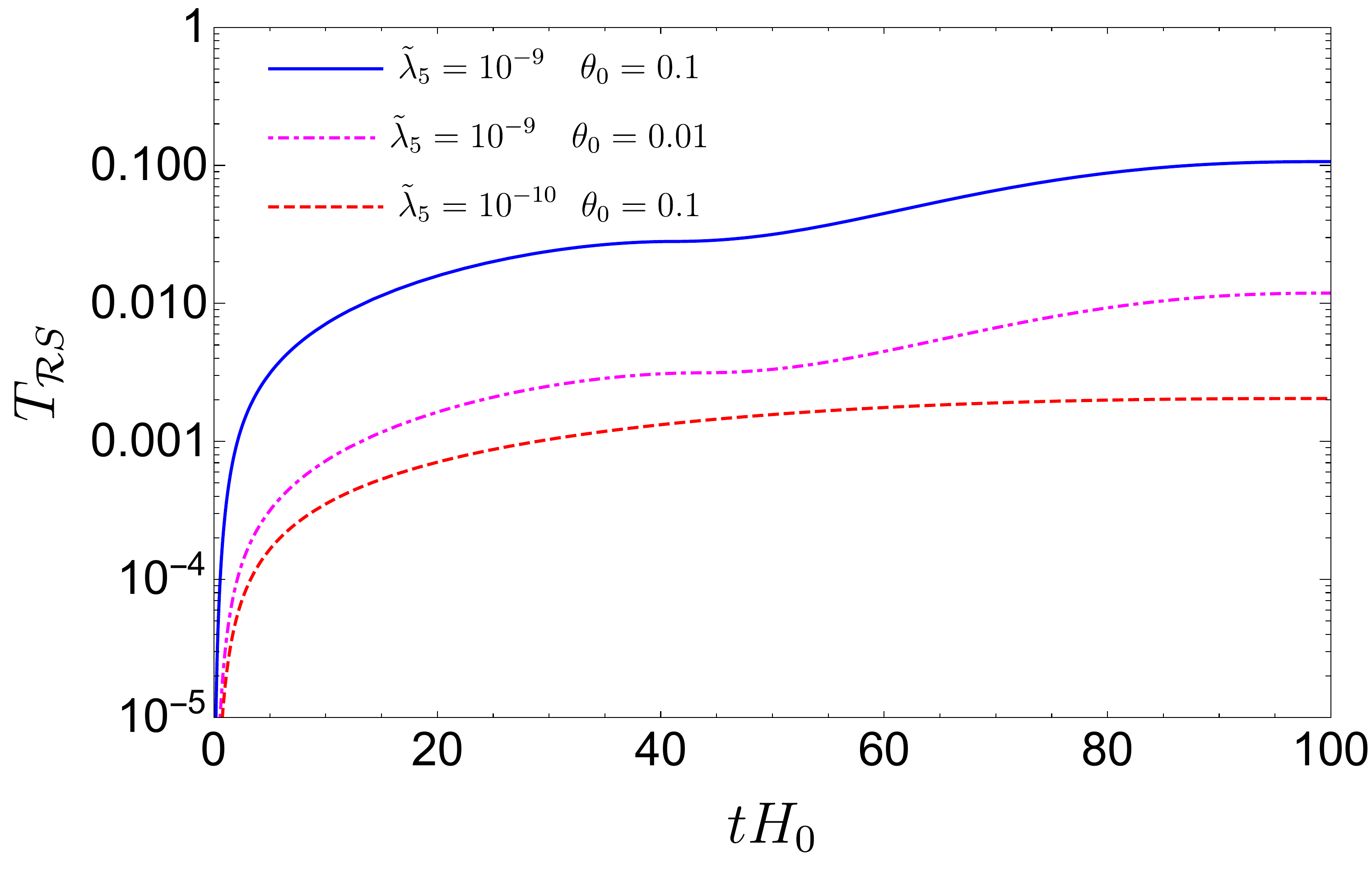}
      	\caption{The evolution of $T_{\mathcal RS}$ during inflation for different input parameters. The input parameters are fixed to $ \xi= 300$ and $\lambda=4.5 \cdot 10^{-5}$, with initial conditions  $\chi_0=6.0 M_p$, $\dot \chi_0=0$,  and $\dot \theta_0=0 $ chosen. We choose $\tilde\mu$ to be sufficiently small to not affect the dynamics.}
      	\label{TRS}
      \end{figure*}

In Fig.~\ref{TRS} we show the evolution of  $T_{\mathcal RS}$, which should be smaller than $0.1$ to avoid the current observational constraint. As shown in the plot, the parameters $\tilde \lambda_5=10^{-9}$ with $\theta_0=0.1$ lead to isocurvature perturbations close to the current sensitivity of CMB observation.  A reduction of $\tilde \lambda_5$ or $\theta_0$ leads to a smaller $T_{\mathcal RS}$. This fact is consistent with the expectation that when $\tilde \lambda_5=0$ or $\theta_0=0$ the isocurvature mode disappears and no lepton asymmetry is generated.

\subsection{Derivation of Inflationary Trajectory in Higgs Portal Inflation}
\label{Inf_tra_dec}

The inflationary context we utilise exhibits a flat direction fixed by the ratio of the two fields $ h $ and $ \Delta^0 $ . Following the derivation in Ref.~[57], we demonstrate the existence of this trajectory. Firstly, consider the field redefinitions of $ h $ and $ \Delta^0 $,
	\begin{equation}
 \chi =  \sqrt{\frac{3}{ 2}} M_p \log \left(1+ \frac{\xi_H h^2}{ M_p^2} +\frac{\xi_\Delta (\Delta^0)^2}{ M_p^2}\  \right) \textrm{~~and~~} \kappa = \frac{h}{ s} ~. 
	\end{equation}
The kinetic terms of the Lagrangian are then given by, 
\begin{align}
{\cal L}_{\rm kin} =& \frac{1}{ 2} \biggl( 1+ \frac{1}{ 6} \frac{\kappa^2 +1}{
	\xi_H \kappa^2 +\xi_\Delta} \biggr)~ (\partial_\mu \chi)^2 + 
\frac{M_p}{\sqrt{6}}  ~\frac{(\xi_\Delta-\xi_H) \kappa }{ (\xi_H \kappa^2 +\xi_\Delta)^2} 
(\partial_\mu \chi) (\partial^\mu \kappa) \nonumber\\
&+
\frac{M_p^2 }{ 2}\frac{ \xi_H^2 \kappa^2 +\xi_\Delta^2 }{ (\xi_H \kappa^2 +\xi_\Delta)^3 }(\partial_\mu \kappa)^2
~. 
\end{align}
We require large non-minimal couplings in our analysis,
$ \xi \equiv \xi_H  +\xi_\Delta \gg 1$, so at leading order in $1/\xi$ the kinetic terms,
\begin{equation}
{\cal L}_{\rm kin}= \frac{1}{ 2 } (\partial_\mu \chi)^2 +
\frac{M_p^2}{ 2} \frac{ \xi_H^2 \kappa^2 +\xi_\Delta^2 }{ (\xi_H \kappa^2 +\xi_\Delta)^3 }(\partial_\mu \kappa)^2 ~. 
\end{equation}
 The large $ \xi $ limit suppresses the mixing term $(\partial_\mu \chi) (\partial^\mu \kappa) $ giving a canonically normalised $ \chi $, while also suppressing the kinetic term of $ \kappa $. There are three key regimes for $ \kappa $, each with corresponding canonically normalized variable $\kappa'$,
\begin{eqnarray}
&& \xi_\Delta \gg \xi_H  ~~{\rm or}~~ \kappa \rightarrow 0 ~~,~~~~ \kappa'= \frac{\kappa}{
	\sqrt{\xi_\Delta}}  ~, \nonumber\\
&& \xi_H \gg \xi_\Delta  ~~{\rm or}~~ \kappa \rightarrow \infty ~~,~~ \kappa'= \frac{1}{
	\sqrt{\xi_H} \kappa}  ~, \nonumber\\
&& \xi_H = \xi_\Delta  ~~,~~~~~~~~~~~~~~~~~~~  \kappa'= \frac{1}{{\sqrt{\xi_H}}} \arctan \kappa ~. 
\end{eqnarray}

Now consider the potential in terms of $ \kappa $,
\begin{equation}
U= \frac{\lambda_H \kappa^4 + \lambda_{h\Delta} \kappa^2 +\lambda_\Delta  }{
	4 (\xi_H \kappa^2 +\xi_\Delta)^2}  M_p^4~,
\end{equation}
for large $ \chi $. This potential has the following minima, dependent upon the chosen coupling relations,
\begin{align}
 (1)&~2 \lambda_H \xi_\Delta - \lambda_{h\Delta} \xi_H >0~,~
2 \lambda_\Delta \xi_H - \lambda_{h\Delta} \xi_\Delta >0~,~~~~\kappa = \sqrt{\frac{ 
	2 \lambda_\Delta \xi_H - \lambda_{h\Delta} \xi_\Delta  }{ 
	2 \lambda_H \xi_\Delta - \lambda_{h\Delta} \xi_H}   }  ~,  \nonumber\\
 (2)&~2 \lambda_H \xi_\Delta - \lambda_{h\Delta} \xi_H >0~,~
2 \lambda_\Delta \xi_H - \lambda_{h\Delta} \xi_\Delta <0~,~~~~\kappa=0  ~,  \nonumber\\
(3)&~2 \lambda_H \xi_\Delta - \lambda_{h\Delta} \xi_H <0~,~
2 \lambda_\Delta \xi_H - \lambda_{h\Delta} \xi_\Delta >0~,~~~~\kappa=\infty  ~, \nonumber\\
 (4)&~2 \lambda_H \xi_\Delta - \lambda_{h\Delta} \xi_H <0~,~
2 \lambda_\Delta \xi_H - \lambda_{h\Delta} \xi_\Delta <0~,~~~~\kappa=0,\infty  ~. 
\end{align}
Case 2 and 3 concern inflationary scenarios dominated by  $ \Delta^0 $ and $ h $, respectively, while the inflaton in scenario 1 is characterised by a mixture of the two scalars.  In scenario 1, the potential has the following minimum,
\begin{equation}
U\Bigl\vert_{\rm min~(1) }= \frac{1}{16} \frac{4 \lambda_\Delta \lambda_H - \lambda_{h\Delta}^2}{\lambda_\Delta \xi_H^2 + \lambda_H \xi_\Delta^2 - \lambda_{h\Delta} \xi_\Delta \xi_H } M_p^4~, 
\label{Umin1}
\end{equation}
which will be related to the Starobinsky mass scale through $\sim\frac{3}{4}m_S^2 M_p^2$ . In cases 2 and 3 we derive the usual single field non minimally coupled inflationary  potential,  $\lambda_\Delta /(4\xi_\Delta^2) M_p^4$ and 
$\lambda_H /(4\xi_H^2) M_p^4$, respectively. 
We require that the numerator  of Eq. (\ref{Umin1}) is positive to ensure that we do not have a negative vacuum energy at large field values. In each case, the canonical field $\kappa'$ obtains a large mass of order  $M_p/ \sqrt{\xi}$. This mass is always greater than the Hubble rate during inflation, so the $ \kappa $ can be integrated out.

\subsection{Possibility of Q-ball Formation}
\label{Qball}
In Baryogenesis scenarios consisting of scalar fields carrying global charges, there exists the possibility of forming Q-balls. It is important to investigate their stability and regime of formation as they can have interesting phenomenological implications. If the  Q-balls are absolutely stable, they can be  a component of the dark matter relic density, and potentially prevent successful Baryogenesis through sequestering the generated asymmetry from the thermal plasma. In our scenario, any Q-balls that are produced will have decay pathways into fermions through the neutrino Yukawa coupling. This means that as long as the decay time of these processes is such that the Q-balls decay before the EWPT, their should be no phenomenological implications of Q-ball formation during the inflationary and reheating epochs.

Firstly, it must be determined whether Q-ball formation is possible in our model. To do this we investigate the effective potential  $U_{\textrm{eff}}(\chi,\omega)=U(\chi)-\frac{1}{2} \omega^2 \chi^2$, and obtain the range of $ \omega $ frequencies for which bounce solutions exist. The upper and lower bounds on the frequency, $\omega+$ and $\omega_-$ are defined as follows,
\begin{equation}
\omega_+^2=U''(0)~, \textrm{~~~and~~~} \omega_-^2=\left. \frac{2U(\chi)}{\chi^2}\right|_{\textrm{min}}~,
\end{equation}
where $ U(\chi) $ is given in Eq. (\ref{Uchi_sm}). From these relations, we obtain the following requirement for Q-ball formation to occur,
\begin{equation}
U''(0)>\left. \frac{2U(\chi)}{\chi^2}\right|_{\textrm{min}}~.
\end{equation}
where we have $m_\Delta \simeq \sqrt{U^{\prime \prime}(0)}$. 

\begin{figure*}[h]
      	\centering
      	\includegraphics[width=0.45\textwidth]{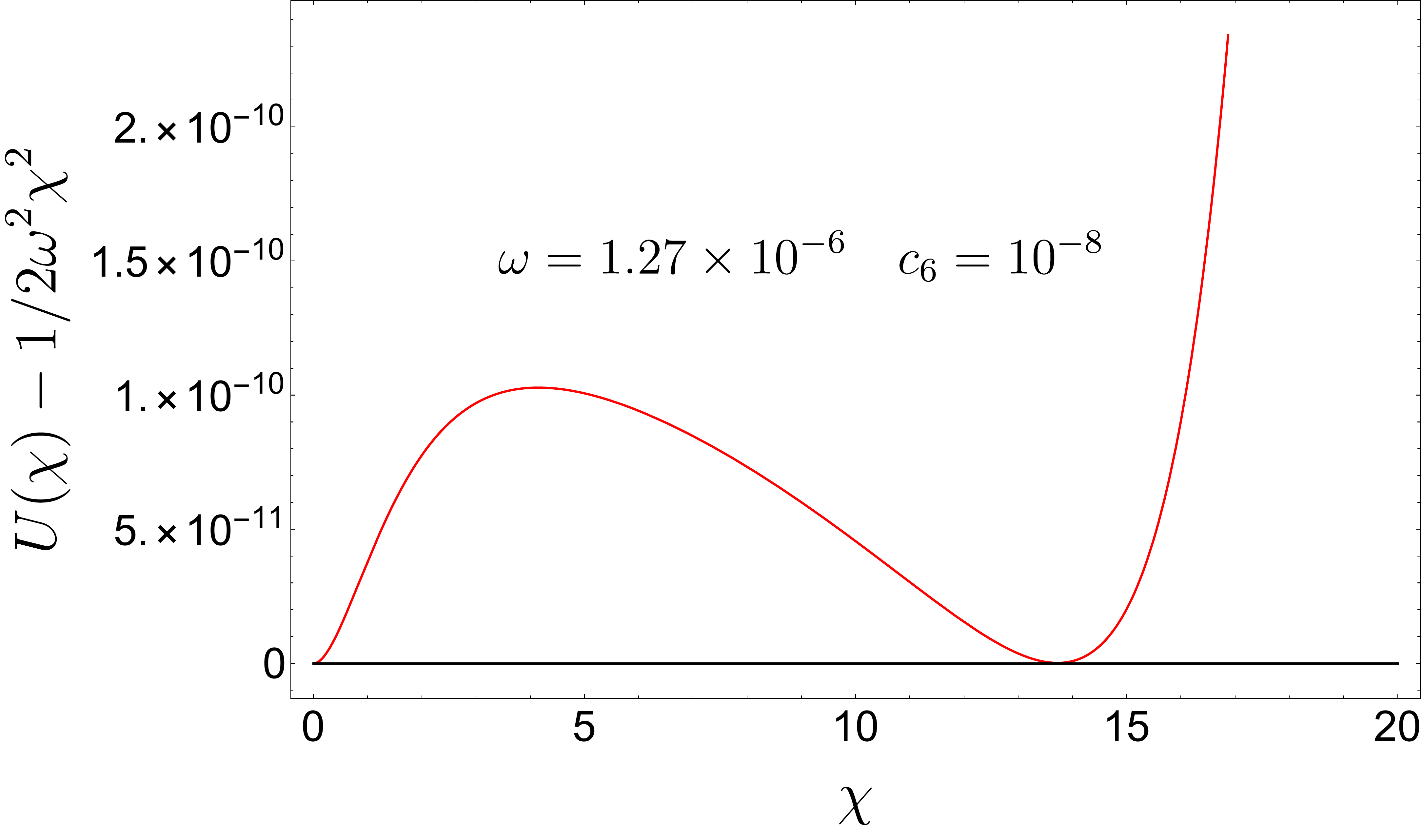}
      	\caption{Potential for bounce solutions. All the parameters are in Planck units.}
      	\label{bounce}
      \end{figure*}

The $\omega_-$ appears when the potential just obtains two degenerate minima, one is at the origin and the other is at a large $\chi$ field value. Since the potential becomes flat for $\chi\gg M_p$, all $\omega_->0$ will allow Q-ball formation for sufficiently large $\chi$. However, it would be expected that higher dimensional operators in $\phi$ should be present in the potential. In terms of the canonically normalised Einstein frame field $\chi$, these higher order terms are exponentially enhanced by $e^{\frac{n}{\sqrt{6}}\frac{\chi}{M_p}}$, for a $n+4$ dimensional term. A dim-6 operator of the form ${\lambda_6}{\frac{\phi^6}{M_p^2}}$, would increase faster with $\chi$ than the $\omega_-$ dependent term, with the lower bound on $\omega$ dependent upon the choice of ${\lambda_6}$ . Note that  $\lambda_6$ should be sufficiently small as to not distort the inflationary dynamics. %If we depict the dependence of $ U_{\textrm{eff}}(\chi,\omega) $ on $ \omega $ for large field values, 
An example is depicted in Figure \ref{bounce}, for which $\omega_{-} \simeq 1.27 \times 10^{-6} M_{p} $ is the lower bound for the choice $\lambda_6=10^{-8}$.  This value of $\omega_{-}$ is much greater than the range of masses $m_\Delta$ that we consider. An approximate relation for when Q-Ball formation occurs can be found between the $\lambda_6$ coupling and the frequency $\omega_-$, as follows,
\begin{equation}
\lambda_6\simeq 430 \frac{\omega_-}{M_p} e^{-\frac{m_S}{\omega_-}}
\end{equation}
From this relation, we see the exponential suppression that is required to allow Q-ball formation for small frequencies. If we consider $\omega_- \sim m_\Delta\sim 1$ TeV, then  the necessary $\lambda_6$ must be incredibly tiny, less than $\sim 2\cdot 10^{-13} e^{-3 \times 10^{10}}$. Thus, we can conclude that the formation of stable Q-balls does not occur in this model.

\section{Behaviour of $\mu$ term}
In this section we give further comment on the trilinear term $\tilde \mu$.  Comparing with the quartic term and dim-5 term, the cubic term becomes increasingly relevant as $\varphi$ decreases. At certain times, the field will not rotate in the phase space, but rather oscillate. Thus, the baryon asymmetry starts to oscillate and we lose the predictability of our model. This is shown in the top-left panel  of Fig.~\ref{nL_mu1}, where the $U(1)_L$ breaking term only includes the $\tilde \mu$ term. 

\begin{figure*}[h]
      	\centering
      	\includegraphics[width=0.45\textwidth]{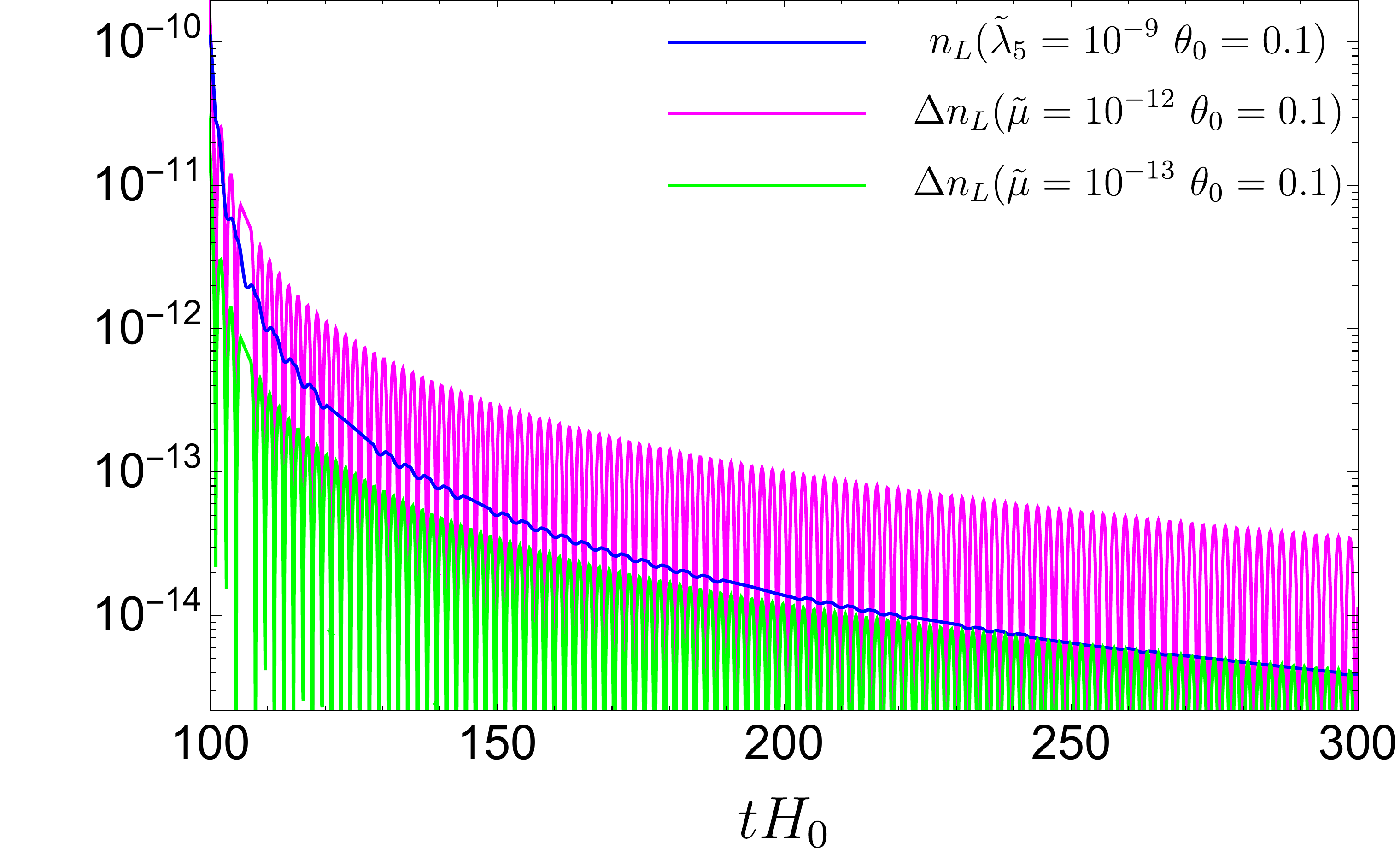}
      	\caption{Illustrating the dependence of the dynamics of $n_L$ on $\tilde{\mu}$ after inflation. The input parameters are fixed to $ \xi= 300$, $\lambda=4.5 \cdot 10^{-5}$ and $\tilde \lambda_5= 10^{-9}$, with initial conditions  $\chi_0=6.0 M_p$, $\dot \chi_0=0$,  and $\dot \theta_0=0 $ chosen. }
      	\label{muterm}
      \end{figure*}

\begin{figure}[t!]
      \begin{subfigure}
      	\centering
      	\includegraphics[width=0.45\textwidth]{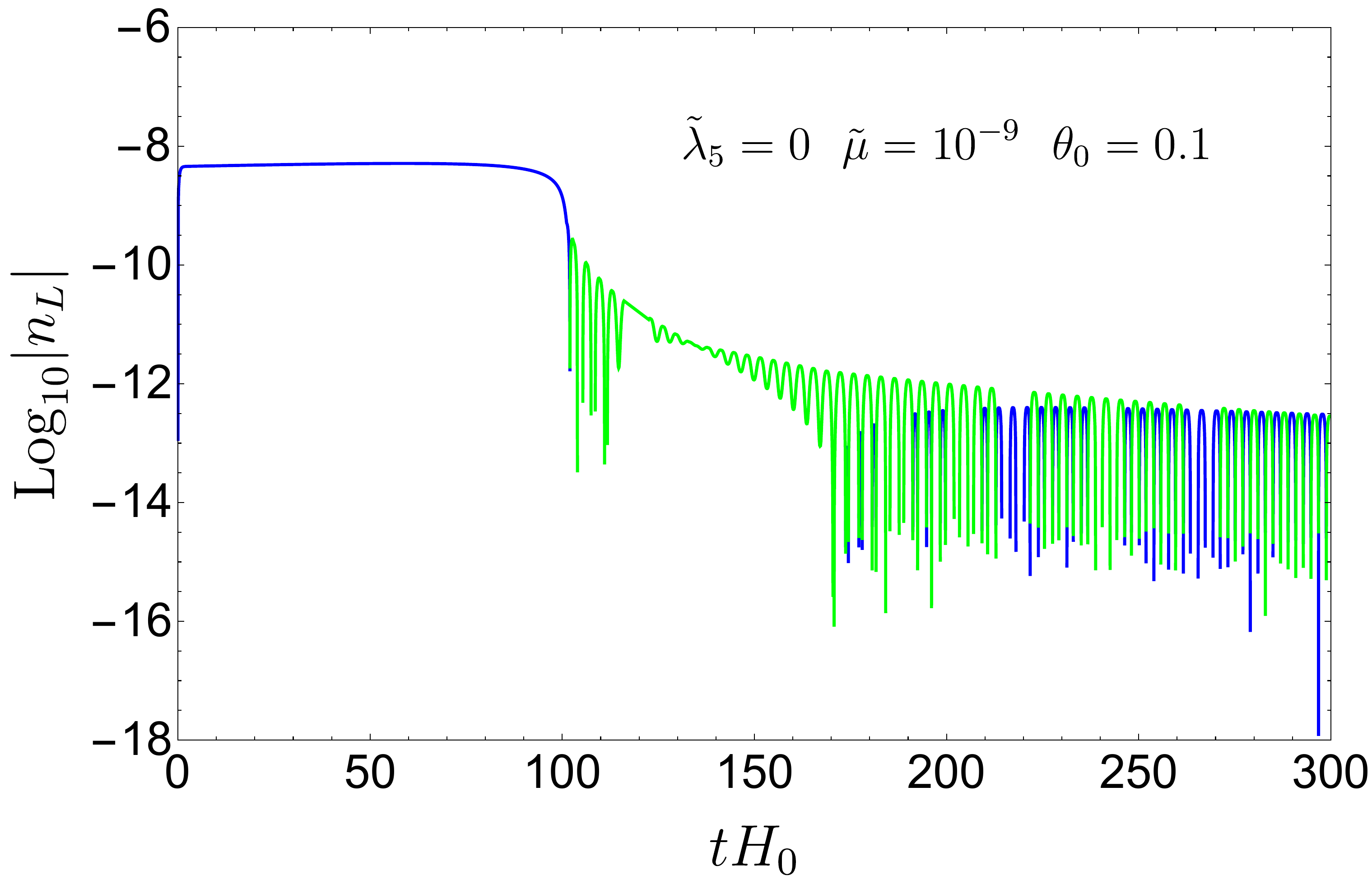}
      	\end{subfigure}
      	\hfill 
      	   \begin{subfigure}
      	   	\centering
      	   	\includegraphics[width=0.45\textwidth]{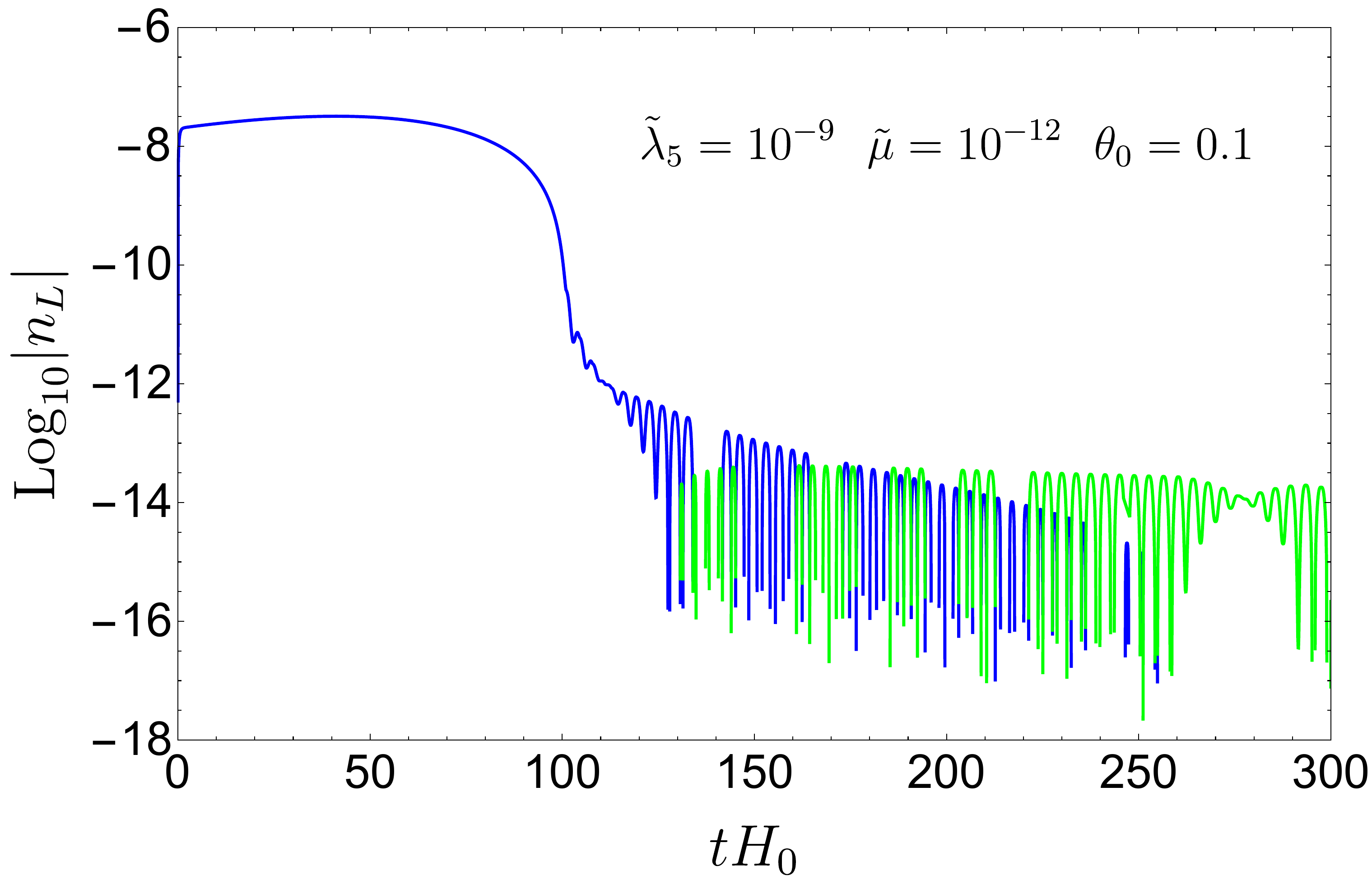}
      	   	\end{subfigure}
      	   	      \begin{subfigure}
      	\centering
      	\includegraphics[width=0.45\textwidth]{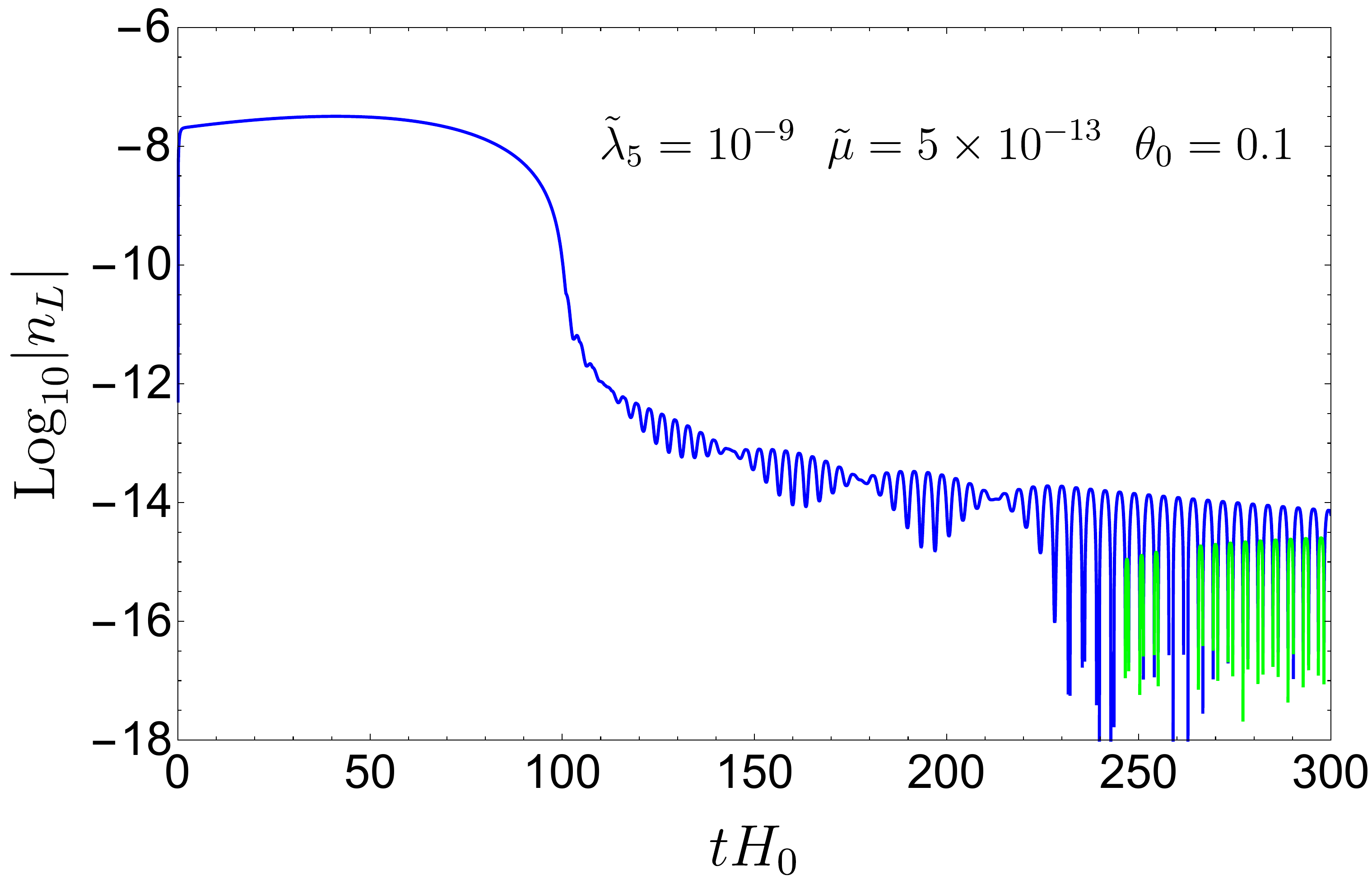}
      	\end{subfigure}
      	\hfill 
      	   \begin{subfigure}
      	   	\centering
      	   	\includegraphics[width=0.45\textwidth]{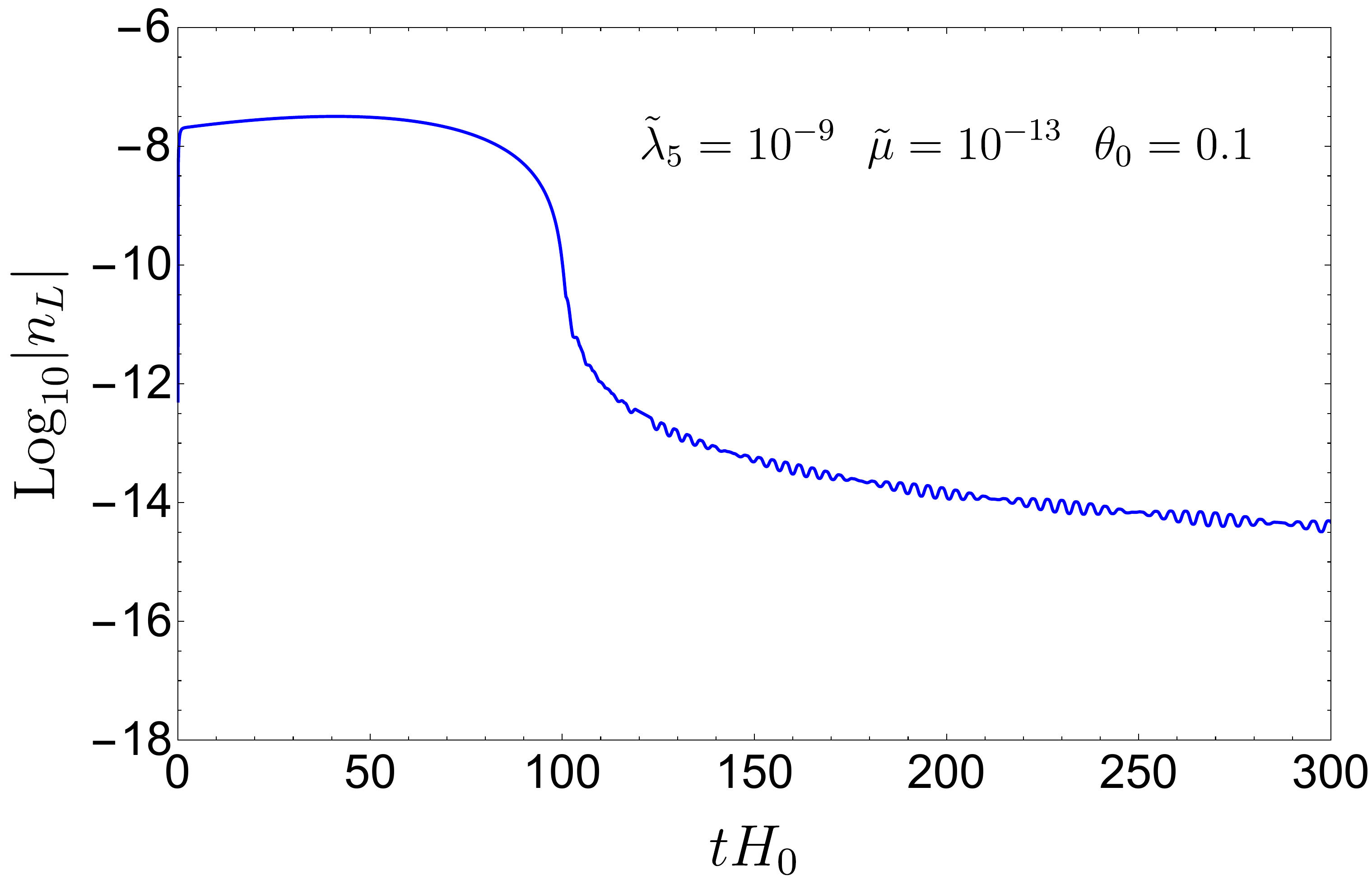}
      	   	\end{subfigure}
      	\caption{  The lepton number density during inflation and and the oscillation epoch for varying cubic coupling term $\tilde{\mu}$, where  $H_0= m_S/2$ .  The blue and green curves denote positive and negative lepton number densities respectively, and  $ n_L $ is given in Planck units. The input parameters are fixed to $ \xi= 300$ and $\lambda=4.5 \cdot 10^{-5}$, with initial conditions  $\chi_0=6.0 M_p$, $\dot \chi_0=0$,  and $\dot \theta_0=0 $ chosen.}
      	\label{nL_mu1}
      \end{figure}

Let us determine the condition that ensures the baryon asymmetry does not oscillate. During the oscillation stage, the lepton number generated from the cubic term within one oscillation time $1/m_S$ can be approximated as, 
\begin{eqnarray}
\Delta n_L\approx \frac{2 Q_L \tilde{\mu} \varphi^3}{m_S}~.
\label{Dbaryon}
\end{eqnarray}
We must ensure that $\Delta n_L \lesssim n_L$ before reheating is completed, hence,
\begin{eqnarray}
{\tilde \mu \lesssim (m_S {n_L}_{\rm reh})/(4 \varphi_{\rm reh}^3)~.}
\label{baryon}
\end{eqnarray}
For the parameter $\tilde \lambda_5=10^{-9}, $ and hence $ {n_L}_\textrm{end}=1.1\times 10^{-11} M_p^{3}$, we depict  $\Delta n_L$ versus $n_L$ in Fig.~\ref{muterm}. From this plot, we find that a numerical value of $\tilde \mu \lesssim 10^{-13} M_p$ for $\theta_0=0.1$ ensures $\Delta n_L \lesssim n_L$. In Fig.~\ref{nL_mu1} we show the evolution of the lepton asymmetry with different $\tilde \mu$. When $\tilde \mu < 10^{-13} M_p $ the lepton asymmetry indeed becomes stable. For the observed baryon asymmetry today, similarly we obtain $\tilde \mu \lesssim 10^{-18} M_p$.

%This assumption is made to allow analytical and numerical analysis of the $\theta$ dynamics and resultant lepton number density. If the parameter $\mu$ is increased above this limit, the lepton number density enters a highly oscillatory regime during the reheating epoch, once it dominates over the $\varphi$ quartic interaction. The dependence of the oscillation amplitude in $n_L$ on $\mu$ can be observed in Figure \ref{nL_mu1}. As $\mu$ increases, the amplitudes are enhanced until they begin to cross into negative values. This means that for $\mu$ values that violate the above constraint we begin to obtain results that are highly oscillatory around zero for the lepton number density. In such cases, it becomes difficult to determine the predicted lepton number density, as it will be highly dependent upon the end point of reheating. Note that the maximum of the oscillations still retains a similar behaviour to that in the small $\mu$ case. 

%During the inflationary stage, the cubic interaction is suppressed relative to the quartic and Dim-5 term due to the non-minimal interaction. The $\tilde{\mu}$ term can be chosen such that it would effect the inflationary trajectory despite this suppression. This behaviour is shown in Figure \ref{nL_mu1}, where increasing $\tilde{\mu}$ leads to significantly different behaviour during the inflationary stage.  For even larger $\tilde{\mu}$ values the inflationary trajectory will no longer be described by a Starobinsky-like cosmology. 

\end{document}